\documentclass[twocolumn,tighten,times]{aastex62} 
\usepackage{newtxmath} 
\usepackage{fix-cm}
\usepackage[OT1]{fontenc}

\received{\today}
\revised{? ?, ?}
\accepted{? ?, ?}
\submitjournal{Nature Astronomy}

\newcommand{\angstrom}{{$\rm \mathring A$}}
\newcommand{\angstroms}{{\rm \mathring A}}
\newcommand{\galex}{{\it GALEX}~}
\newcommand{\MgII}{\hbox{Mg\,{\sc ii}}}
\newcommand{\MgIILambda}{\hbox{Mg\,{\sc ii}}~$\lambda$2798}
\newcommand{\CIII}{\hbox{C\,{\sc iii]}}}
\newcommand{\CIIILambda}{\hbox{C\,{\sc iii]}}~$\lambda$1909}
\newcommand{\CIV}{\hbox{C\,{\sc iv}}}
\newcommand{\CIVLambda}{\hbox{C\,{\sc iv}}~$\lambda$1549}
\newcommand{\Lya}{\hbox{Ly$\alpha$}}
\newcommand{\LyaLambda}{\hbox{Ly$\alpha$}~$\lambda$1216}

\shorttitle{Quasar UV SED}
\shortauthors{Cai et al.}

\begin{document}

\title{A universal average spectral energy distribution for quasars from optical to extreme ultraviolet}

\email{zcai@ustc.edu.cn, jxw@ustc.edu.cn}

\author[0000-0002-4223-2198]{Zhen-Yi Cai}
\affiliation{CAS Key Laboratory for Research in Galaxies and Cosmology, Department of Astronomy, University of Science and Technology of China, Hefei 230026, China}
\affiliation{School of Astronomy and Space Science, University of Science and Technology of China, Hefei 230026, China}

\author[0000-0002-4419-6434]{Jun-Xian Wang}
\affiliation{CAS Key Laboratory for Research in Galaxies and Cosmology, Department of Astronomy, University of Science and Technology of China, Hefei 230026, China}
\affiliation{School of Astronomy and Space Science, University of Science and Technology of China, Hefei 230026, China}


\begin{abstract}

The well-known anti-correlation between the optical/ultraviolet (UV) emission line equivalent widths of active galactic nuclei and the continuum luminosity (the so-called Baldwin effect) is a long-standing puzzle. One common hypothesis is that more luminous sources have softer spectral energy distribution (SED) in the extreme UV (EUV), as revealed by some observational studies. In this work we revisit this issue through cross-matching SDSS quasars with \galex far-UV/near-UV catalogs and correcting the effect of a severe observational bias of significant UV detection incompleteness, i.e., the more luminous in observed-frame optical, the more likely detected in observed-frame UV. We find that, for \galex detected quasars at $1.8 \lesssim z \lesssim 2.2$, the rest-frame mean UV SED ($\sim$ 500 -- 3000~\angstrom) bewilderingly shows no luminosity dependence at $\log[\nu L_\nu(2200~\rm \mathring A)] > 45$ (up to 47.3), contrary to the standard thin disc model predictions and the observed Baldwin effect in this luminosity range. Probably, the universal mean UV SED is the result of a local atomic-originated process, and in fainter quasars stronger disk turbulence launching more clouds is the main origin of the Baldwin effect. After correcting for the absorption of the intergalactic medium, a rest-frame intrinsic mean EUV SED is derived from a sub-sample of bright quasars and is found to be much redder in the EUV than all previous quasar composite spectra, highlighting the significance of properly accounting for the sample incompleteness. Interestingly, the global consistence between our extremely red mean EUV SED and the line-driven wind model again supports an origin of a local physical process.

\end{abstract}



\section{Introduction}\label{sect:intro}

The luminous quasars and more general active galactic nuclei (AGN) are commonly believed to be powered by gas swirling toward the supermassive black hole (BH) inhabiting galactic center \citep[e.g.,][]{Salpeter1964,LyndenBell1969,Shields1978,MalkanSargent1982,Rees1984}. The gas accretion results in an accretion disk surrounding the BH and producing majority of the radiative output. The most extensively studied accretion disk model for quasars is the multi-color blackbody geometrically thin disk model \citep{ShakuraSunyaev1973}, which can approximately interpret the so-called big blue bump feature of quasars in the optical and ultraviolet (UV) wavelengths \citep[e.g.,][]{Shields1978,MalkanSargent1982,Capellupo2015}. 

A general prediction of this simple thin disk model on the spectral shape of continuum radiation, or the spectral energy distribution (SED), gives the well-known blue shape of big blue bump, $f_\nu \propto \nu^{\alpha}$ with spectral index of $\alpha = +1/3$, where $f_\nu$ is the flux density per frequency $\nu$, from the near-infrared (NIR; $\sim 1 - 2~\mu$m) to the optical ($\sim 4000~{\rm \mathring  A}~- 1~\mu$m) and even to the UV wavelengths ($\sim 100 - 4000~{\rm \mathring  A}$, for clarify divided into three constituents: the near-UV (NUV) beyond $2000~{\rm \mathring A}$, the extreme UV (EUV) beneath $1000~{\rm \mathring A}$, and the far-UV (FUV) in between). Additionally, the SED has a peak, commonly predicted in the EUV, set by the disk temperature of the inner most regions of accretion disk.
For non-rotating black holes, both the shape of the SED and the location of its UV peak depend only on $\lambda_{\rm Edd}$/$M_{\rm BH}$ \citep{KoratkarBlaes1999}, where $\lambda_{\rm Edd}$ is the Eddington ratio (the ratio of the bolometric luminosity to the Eddington luminosity) and $M_{\rm BH}$ is the BH mass.

Remarkably, a blue shape for several quasars in the NIR with $\alpha_{\rm NIR} = + 0.44 \pm 0.11$, consistent with the thin disk prediction, has been measured using the polarization observations \citep{Kishimoto2008}.
However, more challenges against this simple thin disk model come from observations on shorter wavelengths, exploring the most inner and energetic regions of the accretion disk and indicating more complex physics beyond the simple thin disk \citep{KoratkarBlaes1999,Lusso2015}. 

Disk models generally predict blue SEDs as well as big blue bump peaked in the EUV \citep[e.g.,][]{KoratkarBlaes1999,Lawrence2012}. However, in the optical to UV wavebands, a much redder shape with $\alpha_{\rm UV} < 0$ is generally observed. Moreover, the SED becomes softening/redder with decreasing wavelength \citep{Neugebauer1987,OBrien1988a,CristianiVio1990,Francis1991,Cheng1991,Zheng1997,VandenBerk2001}, together with a prominent break at $\sim 1000$~\angstrom~(\citealt{Zheng1997,Telfer2002a,ShangZH2005,Shull2012,Stevans2014,Lusso2015}; however see \citealt{Scott2004}). 
For example, \citet{Francis1991} report a median spectral index of $\alpha_{\rm NUV} \sim -0.32$, measured between 1450 and 5050 \angstrom~for several hundred quasars \citep[see also][for a compilation of many similar measurements]{VandenBerk2001}. Furthermore, by constructing a composite {\it Hubble Space Telescope} ({\it HST}) spectrum of 101 quasars at $z > 0.33$, \citet{Zheng1997} conclude $\alpha_{\rm FUV} = -0.99 \pm 0.05$ and $\alpha_{\rm EUV} = -1.96 \pm 0.15$, measured between 1050 and 2200 \angstrom~and between 350 and 1050 \angstrom, respectively.
Similarly, the continuum colors of quasars implied by disk models are not confirmed by observations, showing prominent discrepancies once involving the UV wavelengths \citep{Bonning2007}.

More delicately, there are a few discussions on the dependence of the UV spectral index on the UV luminosity, however, none consistence has been achieved so far. 
\citet{Kuhn2001} report bluer optical-UV (1285-5100~\angstrom) and FUV (1285-2200~\angstrom) slopes with higher UV luminosity at 1460 and 1285~\angstrom, respectively. \citet{Davis2007} also find bluer NUV (2200-4000~\angstrom) slope with increasing 2200~\angstrom~luminosity, but no strong correlation is found between the FUV (1450-2200~\angstrom) slope and the same UV luminosity. Many others suggest there is no luminosity dependence for the FUV spectral index \citep{OBrien1988a,Cheng1991,Davis2007,Krawczyk2013,Stevans2014,Ivashchenko2014}. For example, \citet{Cheng1991} find that $\alpha_{\rm FUV}$ (1216-2200~\angstrom) is almost independent of 2200~\angstrom~luminosity over more than four orders of magnitude. 

Although most studies suggest a luminosity-independent FUV slope for quasars, a luminosity-dependent EUV slope is generally preferred but with large controversy, probably owing to uncertainties in corrections against intergalactic absorptions.
Using the {\it Far Ultraviolet Spectroscopic Explorer} ({\it FUSE}) spectroscopic observations on 85 quasars with $z < 0.67$, \citet{Scott2004} uncover a significantly bluer EUV slope of $\alpha_{\rm EUV} = -0.56^{+0.38}_{-0.28}$ than those obtained by \citet{Zheng1997} with $\alpha_{\rm EUV} = -1.96 \pm 0.15$. The fact that the median luminosity of the {\it FUSE} quasars is an order of magnitude fainter than that of the {\it HST} sample has been for a long time an evidence for a luminosity-dependent EUV slope. This statement is also claimed by the SED analysis using larger quasar samples with multi-band photometric data. For example, \citet{Trammell2007} claim the same luminosity-dependence by combining the broadband photometric data of the Sloan Digital Sky Survey (SDSS) Data Release 3 (DR3) and the {\it Galaxy Evolution Explorer} ({\it GALEX}) General Data Release 1 (GR1) for a sample of $\sim 200$ quasars at $1.82 \lesssim z \lesssim 2.16$ and $45.5 < \log[\nu L_\nu(2200~\rm \mathring A)] < 47.3$
\footnote{For quasars with redshift of $1.82 \lesssim z \lesssim 2.16$, the initial range of the optical continuum luminosity of the UV-detected quasars quoted by \citet{Trammell2007} is $46.2 \lesssim \log L_{2200} \lesssim 48.0$ (see their Figure~18). However, after having carefully checked the 2200 \angstrom~as well as the 5100 \angstrom~monochromatic luminosities  and compared to those quoted by \citet{Shen2011}, we find that in \citet{Trammell2007} those monochromatic luminosities have been overestimated by exactly a factor of $(1+z)^{1.5}$ if the same Galactic extinction correction is applied. Therefore, we nominate here a smaller luminosity range for their sample by $\sim 0.7$ dex, corresponding to the middle redshift of $z \sim 2$ for the concerned redshift bin.}.
While considering SDSS Data Release 7 (DR7) and \galex General Data Release 6 (GR6), \citet{Krawczyk2013} improve the statistics using $\sim 30,000$ quasars with broad redshift range (mean $z \sim 1.5$) and $44 < \log[\nu L_\nu(2500~\rm \mathring A)] < 47$.
Nonetheless, few works conclude a luminosity-independent EUV slope. Following the analysis of \citet{Zheng1997} but using more {\it HST} spectra for 184 quasars at $z > 0.33$, \citet{Telfer2002a} find no evidence for the dependence of $\alpha_{\rm EUV}$ on the 1100~\angstrom~UV luminosity for 39 radio-quite quasars, but a marginal evidence for a bluer EUV spectrum with increasing luminosity for 40 radio-loud (RL) objects.
\citet{Stevans2014} also show the EUV spectral index does not present obvious dependence on the 1100~\angstrom~UV luminosity for 157 quasars.

The hypothesis that more luminous quasars have softer SEDs has been widely accepted since, as predicted by the photoionization models where harder EUV continuum yields larger emission line intensities \citep{KrolikKallman1988}, it may be intuitively responsible for the well-known Baldwin effect, claiming an anti-correlation between the equivalent width (EW) of emission lines and the continuum luminosity \citep{Baldwin1977}.
However, all previous quasar samples constructed heterogeneously by either spectroscopy or photometry to have different UV luminosity may be subject to a prominent but correction-difficult selection bias, that is, the more luminous in rest-frame UV, the more likely detected in rest-frame EUV. This observational bias has recently been discussed by \citet{VandenBerk2020}, who properly account for the \galex detection limits and find that the EUV colors of quasars are substantially redder than found previously.

Besides being able to put strong constraints on the physics of the inner disk regions in quasars, the exact shape of SED in the EUV waveband as well as its luminosity dependence have many important physical implications, including photoionization models and the associated Baldwin effect \citep{KrolikKallman1988}, the cosmic Helium reionization \citep{Miralda-EscudeOstriker1990}, the detectability of high-$z$ quasars \citep{PicardJakobsen1993,Francis1993}, the properties of the intergalactic medium \citep[IGM;][]{Prochaska2009,OMeara2013,Prochaska2014}, and the origin of broad absorption line outflows in quasars \citep{He2017,He2019,ZhaoQY2021}.
We therefore revisit the luminosity dependence of the EUV AGN continuum through cross-matching SDSS data release 14 (DR14) quasars at $1.82 \lesssim z \lesssim 2.16$ with \galex NUV/FUV catalogs (i.e., its final data release 6/7; GR6/7). We are able to construct in Section~\ref{sect:data} a high signal-to-noise ratio (SNR) yet large sample of EUV-detected quasars, i.e., 3871 sources with $> 3 \sigma$ rest-frame EUV detection. This unique quasar sample, which is larger in size by more than an order of magnitude and deeper in rest-frame UV luminosity by 0.5 dex than those of \citet{Trammell2007}, enables us to properly correct the effect of the aforementioned observational bias as discussed in Section~\ref{sect:analyses}.
Contrary to the simple thin disc model prediction and the common explanation for the Baldwin effect, we reveal in Section~\ref{sect:results_discussion} that the mean UV SED ($\sim$ 500~\angstrom~-- 3000~\angstrom~in the rest-frame) bewilderingly shows no luminosity dependence at $\log[\nu L_\nu(2200~\rm \mathring A)] > 45$ up to 47.3. Meanwhile, after correcting for the intergalactic absorptions, an extremely red intrinsic mean EUV SED is derived for a sub-sample of bright quasars by taking the \galex detection limit into account.
Finally, brief conclusions are summarized in Section~\ref{sect:conclusions}.

Throughout this paper we assume $\Omega_{\Lambda} = 0.7$, $\Omega_{m} = 0.3$, and $H_0 = 70~{\rm km~s^{-1}~Mpc^{-1}}$. In the following, the Galactic extinction correction to the observed-frame UV/optical photometry has been applied, but not the intrinsic extinction correction for the quasar host galaxy.

\section{The Data Set}\label{sect:data}

\begin{figure*}[t!]
\centering
\includegraphics[width=0.33\textwidth]{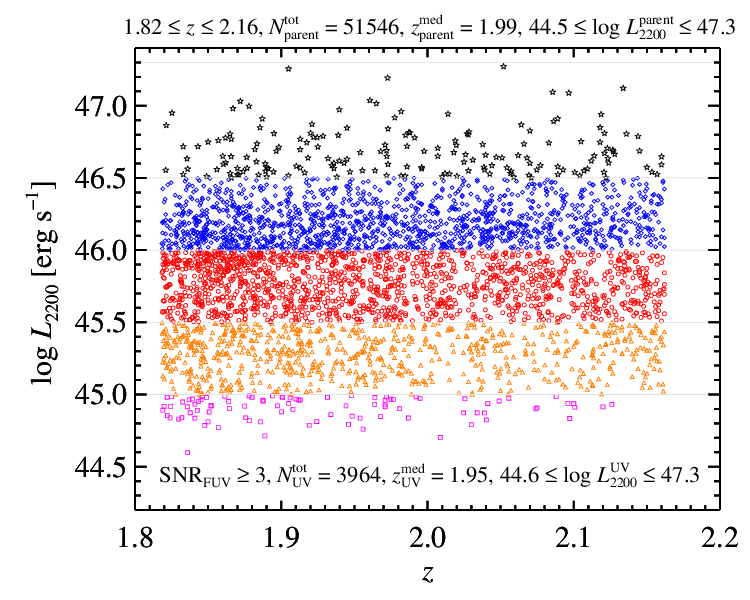}
\includegraphics[width=0.33\textwidth]{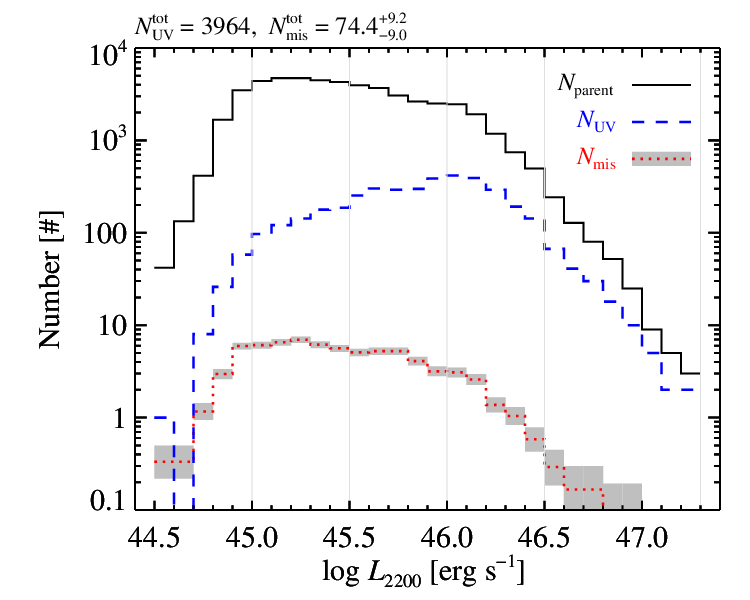}
\includegraphics[width=0.33\textwidth]{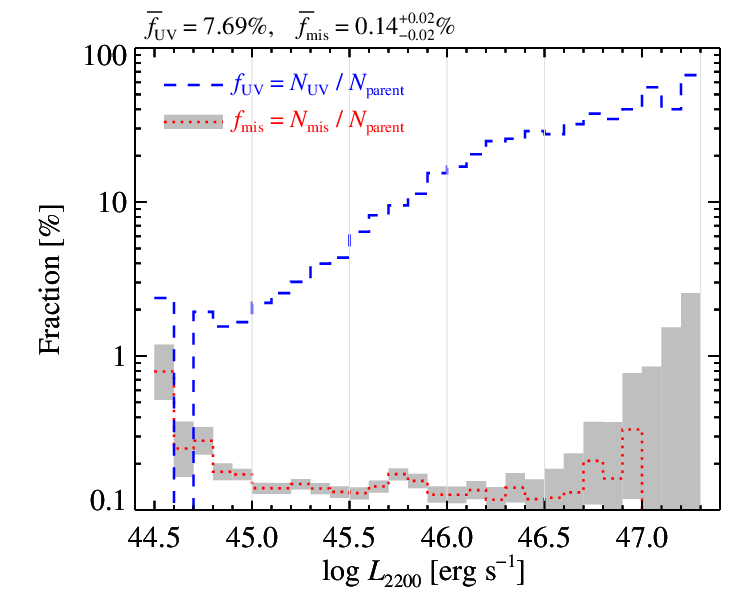}
\caption{
Illustrations of the basic properties of our initial parent quasar sample with $1.82 \lesssim z \lesssim 2.16$ and $44.5 < \log L_{2200} < 47.3$, and the initial UV-detected one with observed-frame SNR$_{\rm FUV} \geqslant 3$. The left panel shows the $\log L_{2200}$ -- redshift distribution for the initial UV-detected quasar sample analyzed in this work. The middle panel shows numbers of the parent quasar (black solid line), the UV-detected quasar (blue dashed line), and the UV mis-matched quasar (red dotted line with gray region for 1$\sigma$ Poisson error) as a function of $\log L_{2200}$. The right panel shows fractions of UV-detected quasar (blue dashed line) and mis-matched quasar (red dotted line with gray region for 1$\sigma$ Poisson error). The parent quasar sample is divided into five independent luminosity bins with bin size of 0.5 dex, except the brightest bin with 0.8 dex (see the distinct colors/symbols in the left panel or the luminosity ranges separated by those vertical light-gray lines in both the middle and right panels).
}\label{fig:sample_selection}
\end{figure*}

The combination of SDSS and \galex observations benefits exploring the luminosity dependence of the UV/optical SED of quasars.
Considering the SDSS DR3 \citep[][]{Abazajian2005} Quasar Catalog \citep[46,420 quasars;][]{Schneider2005}, \citet{Trammell2007} firstly construct a photometric sample of 6371 quasars \footnote{\url{http://vizier.u-strasbg.fr/viz-bin/VizieR?-source=J/AJ/133/1780}}, covered by tiles of the \galex GR1 \citep[][]{Morrissey2005,Martin2005}. Comparing the median SEDs in three luminosity bins for quasars at $1.82 \lesssim z \lesssim 2.16$ (so the rest-frame UV wavelength extending to $\sim 500$ \angstrom) and with $45.5 \lesssim \log L_{2200} \lesssim 47.3$, they find that the median SEDs become bluer (harder) at EUV wavelengths for quasars with lower UV luminosity.
Here $L_{2200} \equiv \nu L_\nu (2200~\rm \mathring A)$ defines the monochromatic luminosity at rest-frame 2200 \angstrom, which has been suggested to be located in a possible emission line free window of the quasar continuum \citep{Francis1991,Cheng1991}.

To further investigate the luminosity dependence of the UV SED of quasars and attempt to correct the aforementioned observational bias, we combine the SDSS DR14 \citep{Abolfathi2018} Quasar Catalog \citep[DR14Q with 526,356 quasars \footnote{\url{https://data.sdss.org/sas/dr14/eboss/qso/DR14Q/DR14Q\_v4\_4.fits}};][]{Paris2018} and the legacy \galex GR6/7 (with 45,195 tiles \footnote{\url{https://galex.stsci.edu/GR6/}}).

\subsection{SDSS quasars}

In order to alleviate as much as possible various observational effects (e.g., the broad emission lines, Lyman absorption, and redshift dependence) that can significantly alter the quasar colors and bias the median SEDs, \citet{Trammell2007} consider three redshift bins and find that there is a prominent luminosity dependence of the EUV portion of quasar SED, detected only in their high-$z$ quasar sample at $1.82 \lesssim z \lesssim 2.16$. Since we are interested in the same EUV portion, we will consider the 66,722 quasars from the SDSS DR14Q within the same redshift range.

For these concerned quasars, we find that in the initial SDSS DR14Q catalog there are $\sim 14\%$ sources with positive magnitude but unphysical flux density (i.e., with a ``-9999'' entry), therefore we update their flux densities according to their positive Asinh magnitudes. Then, we restrict to the 66,432 quasars with positive flux densities and errors in all five SDSS bands.

Again, there are $\sim 14\%$ of the remaining 66,432 quasars with unphysical Galactic extinction \citep[i.e., with a ``-9999'' entry; see also][]{VandenBerk2020}.
For each quasar, we retrieve its $E(B-V)_{\rm SFD}$ from the \citet{Schlegel1998} dust map but assume an $R_V = 3.1$ \citep{Fitzpatrick1999} reddening law, following \citet{SchlaflyFinkbeiner2011}. The adopted ratios of the extinction $A_\lambda$ in the five SDSS bands to the reddening $E(B-V)_{\rm SFD}$ are $A_\lambda/E(B-V)_{\rm SFD} = 4.239$, 3.303, 2.285, 1.698, and 1.263 for SDSS $u$-, $g$-, $r$-, $i$-, and $z$-bands, respectively \citep{SchlaflyFinkbeiner2011}. These values of $A_\lambda/E(B-V)_{\rm SFD}$ are very close to those tabulated in Table~2 of \citet{Yuan2013} for the same \citet{Fitzpatrick1999} reddening law with $R_V = 3.1$.

\subsection{\galex counterparts}

While identifying the \galex counterparts to the SDSS quasars in the \galex imaging survey, we firstly require that, in order to avoid extra scatter on the rest-frame EUV slope once separately searching for the NUV or FUV counterparts, each \galex tile must simultaneously has positive effective exposure times in both NUV and FUV, i.e., $t_{\rm NUV} > 1$ s and $t_{\rm FUV} > 1$ s, resulting in 34,280 ($\simeq 75.8\%$) \galex tiles.

To assess how many SDSS quasars at $1.82 \lesssim z \lesssim 2.16$ are covered by these \galex tiles, we further require that positions of the SDSS quasars are within the central 0.5 degree radius of the field-of-view (FOV) of any one \galex tile, i.e., FOV offset $\leqslant 0.5^\circ$ \citep[][]{Bianchi2014,Bianchi2017}. 
Of the 66,432 SDSS quasars and the 34,280 \galex tiles, we find that 51,587 SDSS quasars are covered by 12,581 \galex tiles. 
Note that \citet{Bianchi2014} introduce the restriction of FOV offset $\leqslant 0.5^\circ$ to avoid poor \galex photometry/astrometry or other artifacts near the edge of the \galex FOV. If the whole \galex FOV is adopted, the number of SDSS quasars covered by \galex tiles would increase by $\sim 12\%$, but properties of the UV detections (discussed below) are nearly the same and all our conclusions are not altered.

Finally, we search for the \galex counterparts to the 51,587 SDSS quasars with a SDSS/\galex offset smaller than 2.6\arcsec~as suggested by \citet{Trammell2007} and find \footnote{\url{https://galex.stsci.edu/casjobs/default.aspx}} 27,281 \galex counterparts with simultaneously positive exposure times in both NUV and FUV. Note part of these counterparts may only have detected NUV or FUV emission. Of the 51,587 SDSS quasars, 15,781 ($\simeq 30.6 \%$) have a unique counterpart and 4959 ($\simeq 9.6 \%$) have more than one counterpart, while the remaining ($\simeq 59.8 \%$) do not have a counterpart.

To pin down a unique \galex counterpart for the 4959 SDSS quasars with multiple \galex counterparts, three criteria have been applied. 
First, the \galex counterpart(s) with longest FUV exposure time is (are) selected, by which 4792 quasars settle down directly. 
Second, even considering the longest FUV exposure, there are 139 quasars with two \galex counterparts and for each counterpart only either NUV or FUV flux density is positive. Since after carefully examining they are all identified in the same \galex tile with identical NUV and FUV exposure times but slightly different coordinates retrieved at NUV and FUV, we merge their NUV and FUV detections as the unique counterpart.
Third, the remaining 28 quasars have two \galex counterparts identified in two different \galex tiles. In this case, we select the one with the smallest FOV offset as the unique counterpart.

For correcting the Galactic extinction in the \galex bands, since they are unavailable in \citet{SchlaflyFinkbeiner2011}, we adopt $A_\lambda/E(B-V)_{\rm SFD} = 6.783$ and 6.620 for the \galex FUV and NUV bandpasses, respectively, from \citet{Yuan2013} for the same \citet{Fitzpatrick1999} reddening law with $R_V = 3.1$.

\subsection{Final quasar sample}\label{sect:sample_summary}

We find 51,587 SDSS quasars at $1.82 \lesssim z \lesssim 2.16$ well covered by \galex imaging tiles. Out of them, 20,371 ($\simeq 39.5\%$) have NUV detections, but only 5990 ($\simeq 11.6\%$) have FUV detections, while 5621 ($\simeq 10.9\%$) have both NUV and FUV detections (hereafter, coined as the UV-detected quasars).
The \galex FUV detection, corresponding to the rest-frame EUV, is crucial for our analysis since both previous photometric and spectroscopic studies (see Section~\ref{sect:intro}) claim more significant luminosity dependence of mean SEDs and composite spectra at shorter EUV wavelengths.

After correcting for the Galactic extinction, we derive the intrinsic monochromatic luminosity at rest-frame 2200~\angstrom, $\log L_{2200}$, for each source by linearly interpolating its SED in $\log(\nu L_\nu)$ versus $\log(\nu)$ space. For the whole 51,587 SDSS quasars, $\log L_{2200}$ covers a very broad range from $\simeq 39.2$ to $\simeq 47.5$, while it is from $\simeq 44.5$ to $\simeq 47.3$ for the UV-detected quasars. There are 39 and 2 SDSS quasars beneath $\log L_{2200} \simeq 44.5$ and beyond $\log L_{2200} \simeq 47.3$, respectively, and none of them is UV-detected. The two most luminous quasars are indeed detected in NUV but not in FUV, confirmed by the \citet{VandenBerk2020} catalog. 
Therefore, to explore the luminosity dependence of the EUV SED of quasars which requires UV detections, we end up with an initial parent SDSS quasar sample having $N^{\rm tot}_{\rm parent} = 51,546$ sources in total with median redshift $z_{\rm parent}^{\rm med} \sim 2$ and $\log L_{2200}$ between $\simeq 44.5$ and $\simeq 47.3$. 

For those UV-detected quasars, 4, 111, and 1657 of them have a SNR in the observed-frame FUV band (SNR$_{\rm FUV}$) less than one, two, and three, respectively. 
Throughout the SNR in $j$-band, SNR$_j$, is estimated according to SNR$_j = f / e_f = 2.5 / \ln(10) e_m$, where $f$, $e_f$, and $e_m$ are the flux density, the error of the flux density, and the error of magnitude at the effective wavelength of $j$-band, respectively.
Since a majority ($\simeq 91.7\%$) of the UV-detected quasars have larger SNR in the observed-frame NUV band than that in FUV band, we prefer using SNR$_{\rm FUV}$ to indicate the quality of \galex photometry.
To increase the confidence on analyzing the UV SED of quasars, we would finally consider an initial UV-detected quasar sample having $N^{\rm tot}_{\rm UV} = 3964$ sources in total with SNR$_{\rm FUV} \geqslant 3$, whose median redshift $z_{\rm UV}^{\rm med}$ and dynamical range of $\log L_{2200}$ are almost the same as those of the parent sample. 

The left panel of Figure~\ref{fig:sample_selection} illustrates $\log L_{2200}$ versus redshift for the initial UV-detected quasar sample (distinct colored symbols for five different luminosity bins), the middle panel shows the $\log L_{2200}$ distributions for both the initial parent SDSS quasar sample (the black solid line) and the initial UV-detected quasar sample (the blue dashed line), and the right panel presents the UV detection fractions, $f_{\rm UV} \equiv N_{\rm UV}/N_{\rm parent}$, as a function of $\log L_{2200}$ (the blue dashed line), where $N_{\rm parent}$ and $N_{\rm UV}$ are numbers of the initial parent SDSS and UV-detected quasars in each $\log L_{2200}$ bin.

As discussed by \citet{Trammell2007}, there is a small probability of a UV source to be coincident with an SDSS quasar. In Figure~\ref{fig:sample_selection}, we also estimate the mis-matched fraction, $f_{\rm mis} \equiv N_{\rm mis}/N_{\rm parent}$, as a function of $\log L_{2200}$ (the red dotted line superimposed on the gray region), where $N_{\rm mis}$ is the number of mis-matched quasars in the same $\log L_{2200}$ bin as $N_{\rm parent}$. 
Using the whole 51,587 SDSS quasars to estimate the mis-matched fraction, we first shift their right ascension, declination, or both by $\pm$30\arcsec, $\pm$60\arcsec, $\pm$90\arcsec, and $\pm$120\arcsec, respectively, and then cross-match with the \galex catalog following the same aforementioned procedure before calculating the number of the mis-matched UV-detected quasars as a function of $\log L_{2200}$. In total, we repeat the estimations 24 times and illustrate in Figure~\ref{fig:sample_selection} the average mis-matched number (middle panel) and average mis-matched fraction (right panel) with 1$\sigma$ Poisson error. 
Compared to the initial UV-detected quasar sample, the number of mis-matched quasars is about $74.4$ and so the probability of a UV source to be coincident with an SDSS quasar is about 1.8\%, which is comparable to that quoted by \citet{Trammell2007}. As shown in the right panel of Figure~\ref{fig:sample_selection}, the mis-matched fraction is nearly independent of luminosity as expected. Therefore, except for $\log L_{2200} < 45$, the UV mis-matched bias is negligible.

As shown in Figure \ref{fig:sample_selection}, numbers of the initial parent SDSS quasars increase with decreasing luminosity except the lowest luminous bin. This indicates the initial parent SDSS quasar sample in the lowest luminosity bin of $44.5 < \log L_{2200} < 45$ is highly incomplete. 
The UV detection fraction in this bin is also very low ($\simeq$ 1.62\%) and it is heavily subject to the mis-matching problem. 
Furthermore, to correct against observational biases as we should introduce in Section~\ref{sect:perform_corrections} by matching the UV detection fraction to this lowest luminosity bin would lead to the loss of the overwhelming majority of UV detections in other luminosity bins.
Therefore, to be more convincing on our results, we drop this lowest luminosity bin and seriously consider quasars brighter than $\log L_{2200} = 45$ in the following.

In sum, the final parent SDSS quasar sample have 45,792 sources at $1.82 \lesssim z \lesssim 2.16$ and with $45 < \log L_{2200} < 47.3$, in which 3871 ($\simeq 8.5\%$
\footnote{Here, our UV detection fraction is significantly lower than that (48\%) nominated in the abstract of \citet{Trammell2007} for the whole 6371 SDSS DR3 quasars using a SDSS/\galex offset of 7\arcsec. However, the comparison is not so straightforward due to the very different selection criteria, including range differences in redshift, luminosity, SDSS/\galex offset, FOV offset, and SNR$_{\rm FUV}$. 
In the \citet{Trammell2007} sample, there are 965 quasars at $1.82 \lesssim z \lesssim 2.16$ and with $\log L_{2200} > 45.5$, 199 ($\simeq 20.6\%$) of which are UV-detected with SDSS/\galex offset smaller than 2.6\arcsec, FOV offset $\leqslant 0.6^\circ$, and SNR$_{\rm FUV} \geqslant 1$. Based on these UV-detected quasars, they conclude the luminosity dependence of quasar EUV SED.
Once considering the same selection criteria (i.e., $1.82 \lesssim z \lesssim 2.16$ and $45.5 < \log L_{2200} < 47.3$ for selecting the parent SDSS quasars as \citealt{Trammell2007}, while SDSS/\galex offset smaller than 2.6\arcsec, FOV offset $\leqslant 0.5^\circ$, and SNR$_{\rm FUV} \geqslant 3$ for selecting the UV-detected ones as ours, stricter than \citealt{Trammell2007}), the UV detection fraction of our quasar sample, i.e., $\simeq 13.6\%$ for 3145 UV-detected in 23,199 SDSS quasars, is as expected higher (due to deeper \galex images) than that of \citet{Trammell2007}, where 29 ($\simeq 3.0\%$) of 965 SDSS quasars are UV-detected.
}
; hereafter, the final UV-detected quasar sample) are UV-detected in both \galex bands with SNR$_{\rm FUV} \geqslant 3$. Of these UV-detected quasars, 2828 ($\simeq 73.1\%$), 1039 ($\simeq 26.8\%$), and 4 ($\simeq 0.1\%$) have $t_{\rm NUV} \simeq t_{\rm FUV}$ ($\leqslant 1\%$ difference), $t_{\rm NUV} > t_{\rm FUV}$, and $t_{\rm NUV} < t_{\rm FUV}$, respectively.

\section{Analyses}\label{sect:analyses}

\subsection{The median-normalized SEDs}

\begin{figure*}[t!]
\centering
\includegraphics[width=0.9\textwidth]{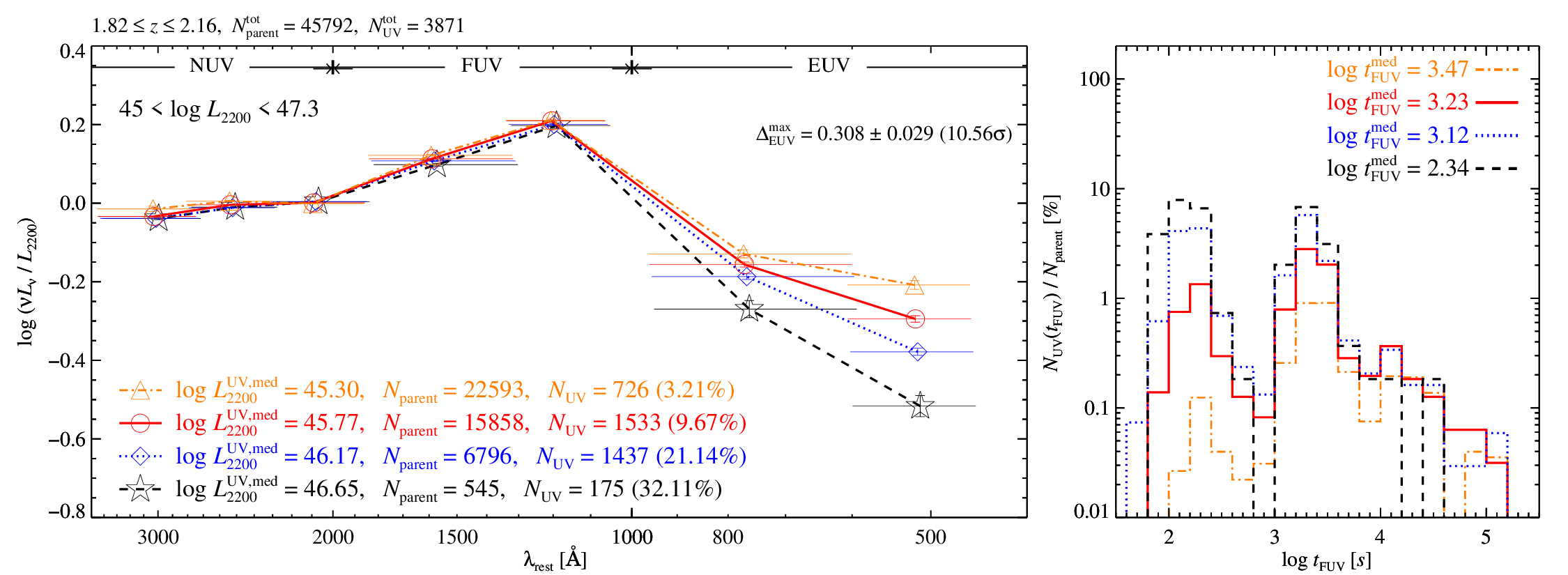}
\includegraphics[width=0.9\textwidth]{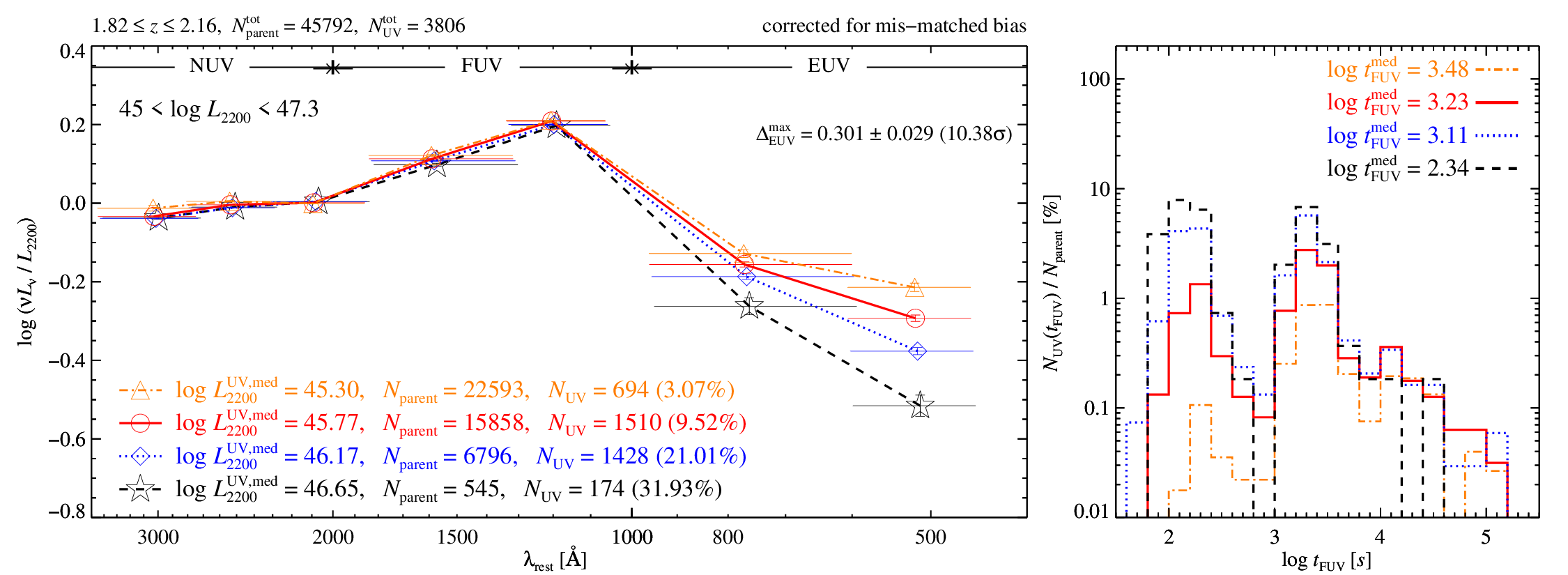}
\includegraphics[width=0.9\textwidth]{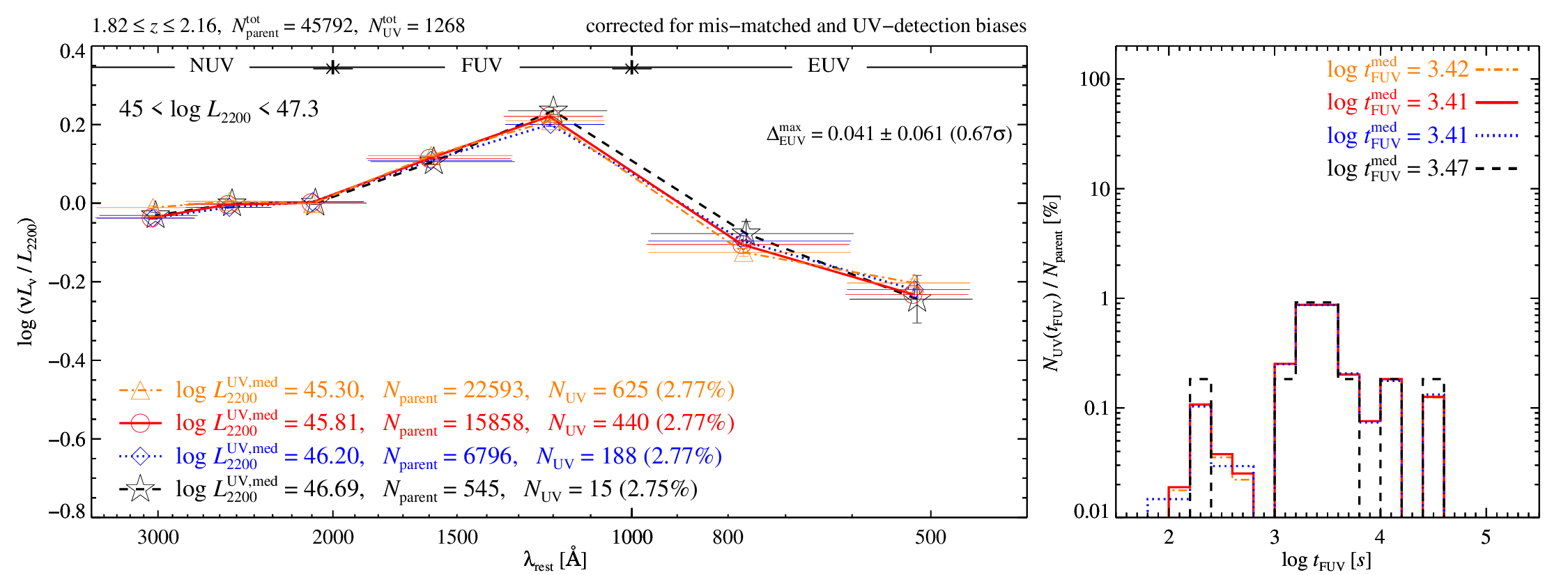}
\caption{
Top panels: the rest-frame NUV to EUV median-normalized SEDs (left panel) and the corresponding relative distributions of \galex FUV exposure times (right panel) for the $N^{\rm tot}_{\rm UV}$ UV-detected quasars at $1.82 \lesssim z \lesssim 2.16$ and with $45 < \log L_{2200} < 47.3$, selected from the $N^{\rm tot}_{\rm parent}$ parent SDSS quasars and separated into four different bins of UV luminosity at 2200~\angstrom. 
In the left panel, with increasing UV luminosity, the median-normalized SEDs are illustrated as the triangles, circles, diamonds, and stars linked by the orange dot-dashed ($45 < \log L_{2200} < 45.5$), red solid ($45.5 < \log L_{2200} < 46$), blue dotted ($46 < \log L_{2200} < 46.5$), and black dashed lines ($46.5 < \log L_{2200} < 47.3$), respectively.  The vertical error bars indicate 1$\sigma$ statistical errors, while the horizontal bars indicate the SDSS and \galex bandpasses.
In each luminosity bin, the legend nominates from left to right the median 2200~\angstrom~luminosity of the UV-detected quasars, $\log L_{2200}^{\rm UV,med}$, the number of parent SDSS quasars, $N_{\rm parent}$, and the number of UV-detected quasars, $N_{\rm UV}$, with the UV detection fraction of $N_{\rm UV}/N_{\rm parent}$. 
A quantity, $\Delta^{\rm max}_{\rm EUV}$, is used to crudely indicate the maximal difference and its significance among these SEDs in the rest-frame shortest EUV wavelength (corresponding to the observed-frame \galex FUV band).
In the right panel, the legends include the median \galex FUV exposure times, $\log t_{\rm FUV}^{\rm med}$, for distinct luminosity bins, clearly unveiling typical smaller \galex FUV exposure times for brighter quasars.
Middle panels: same as the top panels but for only correcting against the mis-matched bias (Section~\ref{sect:mismatch}).
Bottom panels: same as the top panels bur for correcting against both the mis-matched and UV detection incompleteness biases (Section~\ref{sect:UV_detected_bias}).
}\label{fig:SED_bin}
\end{figure*}

For the UV-detected quasar sample, we bin them according to their rest-frame 2200~\angstrom~luminosity (following \citealt{Trammell2007}), and construct rest-frame median-normalized SEDs (throughout also named as mean SEDs) utilizing SDSS and \galex photometry.
The top panels of Figure~\ref{fig:SED_bin} illustrate the median-normalized SEDs (left panel) as well as the distributions of \galex FUV exposure times (right panel) for four different luminosity bins, ranging from $\log L_{2200} = 45$ to 47.3 with bin size of 0.5 dex, except 0.8 dex for the brightest luminosity bin in order to include a few bright but sparse sources. 
The observed-frame effective wavelengths of the SDSS $u$-, $g$-, $r$-, $i$-, and $z$-bands are 3551, 4686, 6166, 7480, and 8932~\angstrom~\citep{Stoughton2002}, respectively, while those of the \galex FUV and NUV channels are 1528 and 2271~\angstrom~\citep{Morrissey2005}, respectively.
For each quasar, after correcting for Galactic extinction, deredshifting the observed-frame effective wavelengths, and normalizing to its own flux density at 2200~\angstrom~by linear interpolation in $\log(\nu 
L_\nu)$--$\log \nu$ space, the median-normalized SED is then obtained (symbols linked by lines in the top-left panel of Figure~\ref{fig:SED_bin}) with 1$\sigma$ statistical error estimated as the standard deviation divided by the square root of the number of quasars (vertical error bars in the top-left panel of Figure~\ref{fig:SED_bin}). 

In the left panels of Figure~\ref{fig:SED_bin}, the horizontal bars indicate the rest-frame SDSS and \galex bandpasses which are deredshifted using the median redshift of sources in each luminosity bin. The observed-frame \galex FUV and NUV bandpasses are 1344-1786~\angstrom~and 1771-2831~\angstrom, respectively, defined by \citet{Morrissey2005} as wavelengths with transmission at least 10\% of the peak. Accordingly, we find bandpasses for SDSS bands: 3136-3943~\angstrom, 3885-5423~\angstrom, 5479-6846~\angstrom, 6789-8275~\angstrom, and 8105-10618~\angstrom~for the SDSS $u$-, $g$-, $r$-, $i$-, and $z$-bands, respectively.

Intuitively, as reported by \citet{Trammell2007}, we confirm that more luminous quasars have redder EUV SEDs. 
Adopting $\Delta^{\rm max}_{\rm EUV}$ to represent the maximal difference of these SEDs in the rest-frame shortest EUV wavelength (corresponding to the observed-frame \galex FUV band), the prominence of luminosity dependence of these SEDs is found at a much higher confidence level than that reported by \citet{Trammell2007}.

\subsection{Observational Biases and Corrections}\label{sect:perform_corrections}

Although the median-normalized SEDs are seemingly found to be luminosity dependent, it must be examined against possible observational biases before discussing their physical implications. Actually, there are two obvious biases that could lead to a bluer EUV SED for low-luminosity quasars. 

\subsubsection{Correction against the mis-matched bias}\label{sect:mismatch}

The first bias is due to chance alignments between SDSS quasars and \galex sources, i.e., when cross-matching SDSS quasars with \galex catalog, there is a probability for SDSS quasars being mis-matched to unrelated \galex sources. 

Following what introduced in Section~\ref{sect:sample_summary}, we estimate in average $\simeq 63.7$ mis-matched quasars for 3871 UV-detected sources within $45 \lesssim \log L_{2200} \lesssim 47.3$. With increasing luminosity, the numbers of UV-detected quasars in each luminosity bin are 726, 1533, 1437, and 175, respectively (see the top-left panel of Figure~\ref{fig:SED_bin}), while the corresponding average numbers of mis-matched quasars, $\bar N_{\rm mis}$, are about $\simeq 31.4$, 22.8, 8.7, and 0.8, respectively. In each luminosity bin, we remove from the UV-detected quasar sample at least the same source number determined by the integral part of the average number of mis-matched quasars, while one more source may be removed at a probability according to the decimal part of the average number of mis-matched quasars. Since the mis-matched sources are expected to have relatively bluer EUV SEDs, we estimate for each UV-detected quasar a ratio of $L_{500}$ to $L_{2200}$, where $L_{500}$ is the monochromatic luminosity at rest-frame 500~\angstrom~by linear interpolation in $\log(\nu 
L_\nu)$--$\log \nu$ space. Assuming that the source with the largest $L_{500}/L_{2200}$, i.e., the bluest EUV SED, has the largest probability of being mis-matched, 
we start rejecting the source with the largest $L_{500}/L_{2200}$ at a probability of $P_{\rm reject} = \bar N_{\rm mis} - N_{\rm rem}$, where $N_{\rm rem}$ is the number of sources having been removed. If $P_{\rm reject} \geqslant 1$, the source with bluest EUV SED is definitely rejected and $N_{\rm rem}$ increases by one. Iteratively, until $0 < P_{\rm reject} < 1$, one more source may be rejected a probability of $P_{\rm reject}$. No more source would be rejected as long as $P_{\rm reject} < 0$. 

As expected by the small numbers of the mis-matched quasars, we confirm that after correcting against the mis-matched bias the resultant median-normalized SEDs among different luminosity bins (see the middle-left panel of Figure~\ref{fig:SED_bin}) are similar to, although with a slightly smaller $\Delta^{\rm max}_{\rm EUV}$ than without any correction (see the top-left panel of Figure~\ref{fig:SED_bin}). 

\begin{figure*}[t!]
\centering
\includegraphics[width=0.9\textwidth]{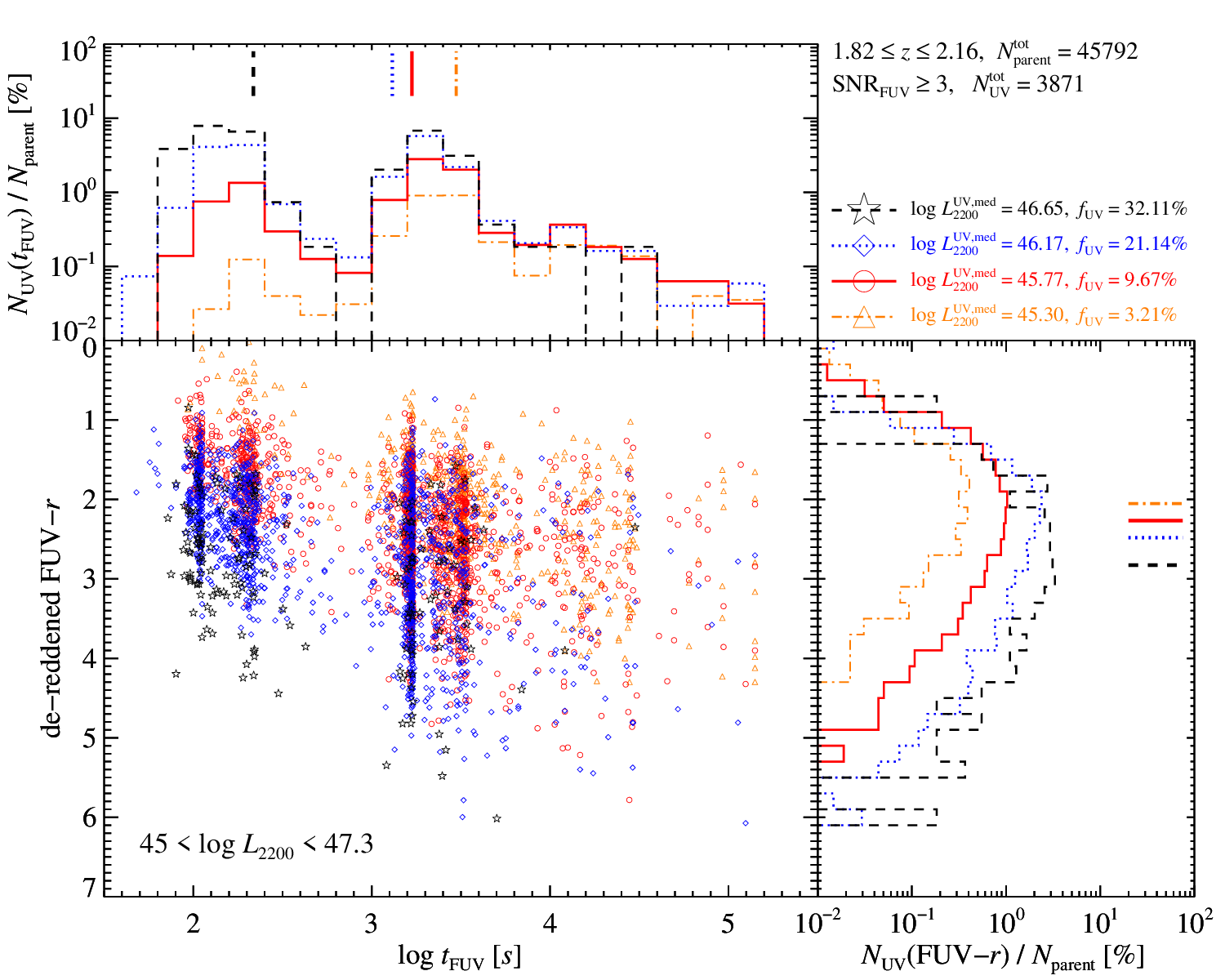}
\caption{
The de-reddened FUV-$r$ color versus the \galex FUV exposure time, and the corresponding distributions, for the UV-detected quasar samples with $45 < \log L_{2200} < 47.3$ and in four luminosity bins, with colored symbols and lines in the same meanings as Figure~\ref{fig:SED_bin}. In the top and left panels, the thick vertical and horizontal bars indicate the medians of $t_{\rm FUV}$ and FUV-$r$ color of the UV-detected quasars in each luminosity bin, respectively. In the main panel, the inhomogeneous distribution of colored symbols intuitively highlights the complexity of \galex observations and its detection limits as a function of $t_{\rm FUV}$ by the missing very red quasars observed shortly, especially for fainter quasars (i.e., a severe observational bias).
}\label{fig:FUV_r_tFUV_dis}
\end{figure*}

\subsubsection{Relative correction against the UV detection incompleteness bias}\label{sect:UV_detected_bias}

The second bias, named as the UV detection incompleteness bias or briefly the UV detection bias, has been presented in the top panels of Figure~\ref{fig:SED_bin} (see also Figure~\ref{fig:FUV_r_tFUV_dis} for more intuitive illustration), where we can find that, from the brightest to faintest luminosity bin, the UV detection fraction of quasars quickly decreases from $\sim 30\%$ to only $\sim 3\%$. 
Because the rest-frame UV less luminous quasars are intrinsically fainter in EUV, only a smaller fraction of them with relatively bright EUV emission can then be detected by \galex. This could naturally result in, as shown in Figure~\ref{fig:FUV_r_tFUV_dis}, that the UV-detected quasars in the less luminous bins have bluer FUV-$r$ colors (at given \galex FUV exposure times). 

This bias has been discussed by \citet{VandenBerk2020} who demonstrate that after properly accounting for the \galex detection limits the EUV colors of quasars are substantially redder than found previously.
The distributions of the effective \galex FUV exposure times for the UV-detected quasars in various luminosity bins, constructed from $\log (t_{\rm FUV}/[s]) = 1$ to 5.2 with bin size of 0.2 dex are also shown in the top panel of Figure~\ref{fig:FUV_r_tFUV_dis}. These distributions are quite different among luminosity bins as indicated directly by their distinct median \galex FUV exposure times, i.e., from the brightest to faintest luminosity bin, the median \galex FUV exposure time of the detections increases from $\sim 200$ s to $\sim 3300$ s. It is caused by that, as the exposure time of \galex tiles spans a broad range, fainter quasars are more likely detected in tiles with longer exposures. 

It may be easy to realize this bias, but really difficult to correct against it.
In analogous analyses of both \citet{Trammell2007} and \citet{Krawczyk2013}, no relevant discussions are presented. Until recently, \citet{VandenBerk2020} tentatively assign to each quasar without \galex detection an upper flux limit determined according to the \galex title with the largest effective exposure time covering that quasar. Note in spite of the significantly redder EUV colors of quasars found by \citet{VandenBerk2020}, they have not explored the luminosity dependence of EUV SEDs. 
As the UV detection fraction of our quasars is rather low, and the EUV emissions of quasars in the \galex bands could be well diverse below the estimated upper flux limit, simply assign flux upper limits to those non-detections would yield a sample dominated by poorly constrained upper limits, hindering further statistical analyses. 
Below we first perform a relative correction relying solely on real \galex detections, to enable direct comparison between luminosity bins. The correction is done through matching the UV detection completeness between luminosity bins. For instance, if the UV detection completeness is A\% in one lower luminosity bin, and a larger B\% in a higher luminosity bin, we drop B\% - A\% of the UV detections in the higher luminosity bin and only keep the rest A\% for comparison with the lower luminosity bin. 
We stress that such relative correction can not yield the intrinsic mean SED for the whole sample in a luminosity bin, but only the mean SED for a portion of EUV brightest sources in each bin.  

However, determining which UV detections to be excluded is not straightforward, because of the highly inhomogeneous \galex FUV exposure times (Figure~\ref{fig:FUV_r_tFUV_dis}). 
In case of if all \galex tiles have uniform depth (i.e., the same exposure time) thus a uniform limiting \galex FUV magnitude,  one may choose to only keep the top brightest UV detections (smaller \galex FUV magnitudes) in the higher luminosity bins, to mimic a lower detection fraction and to ensure a fair comparison with the lower luminosity bin.
Instead, the situation is more complicated considering the exposure time of \galex tiles spans a rather broad range. In this case the limiting magnitude of \galex tiles is no longer uniform, 
thus simply dropping sources with fainter \galex FUV magnitudes in the higher luminosity bin can not precisely mimic the detection incompleteness in the lower luminosity bin. 
The proper approach is to drop sources with smaller SNR$_{\rm FUV}$, as in the lower luminosity bin a uniform cut of SNR$_{\rm FUV} \geqslant 3$ has been applied. 
We stress that because of the spread of the \galex imaging depth and $L_{2200}$ in each luminosity bin, the sources with UV detection or kept after our correction only roughly, but not precisely, represent those with the brightest \galex FUV magnitudes, or the bluest UV SED. For instance, a quasar with fainter \galex FUV magnitude may still be kept/detected because of longer exposure time. Also, a quasar with $L_{2200}$ close to the upper bound of its luminosity bin may be more likely detected, even if it has slightly redder UV SED than other quasars in the same bin.
The \galex tiles come from various programs observing different fields at different depths and with likely variable Galactic extinction and UV source surface density. To minimize potential bias due to these effects, and other possible instrumental effects which are exposure time dependent, we further match the \galex FUV exposure time distributions of the UV detections between luminosity bins. We describe below the detailed procedures. 

In each 0.2 dex bin of $\log t_{\rm FUV}$, we estimate a minimal relative UV detection fraction among all concerned luminosity bins.
Multiplying the parent SDSS quasar number in any luminosity bin by the minimal relative UV detection fraction in a $\log t_{\rm FUV}$ bin and then rounding it, we obtain the number of sources (with top largest SNR$_{\rm FUV}$) to be preserved. In such a way, we obtain matched \galex FUV exposure time distributions, and matched UV detection fractions, for all luminosity bins (see the bottom panels of Figure~\ref{fig:SED_bin}).

Note that when correcting against the mis-matched bias in each luminosity bin, because the expected mis-matched numbers are not round, randomly rejecting the decimal source or not would result in numbers of the UV detections differing by one and slightly fluctuating relative distribution of the \galex FUV exposure times, which in turn somewhat changes the total remaining number of UV-detected sources after further correcting against the UV detection bias. Our conclusions however are not affected by this randomness. Moreover, owing to the marginal effect of correcting against the mis-matched bias, our conclusions are confirmed to be the same if solely correcting against the UV detection incompleteness bias, but we keep both of them for comprehensiveness.

\section{Results and Discussion}\label{sect:results_discussion}

Combing the SDSS DR3 and \galex GR1, \citet{Trammell2007} split their high-$z$ quasar sample at $1.82 \lesssim z \lesssim 2.16$ and with $45.5 < \log L_{2200} < 47.3$ into three equally populated luminosity bins and find that the higher luminosity quasars have a bluer EUV SED. Later on, using SDSS DR7 and \galex GR6, \citet{Krawczyk2013} extend the same conclusion to fainter luminosities, considering three luminosity bins approximately ranging from $\log L_{2500} \sim 44$ to $\sim 47$, but covering quasars within a very broad redshift range approximately ranging from $z \sim 0.5$ to $\sim 2.5$ with a mean redshift of $\sim 1.5$. However, both of them have not tried to correct against observational biases, such as those we have discussed in Section~\ref{sect:perform_corrections}.

In this work, utilizing the SDSS DR14 and \galex GR6/7, we likewise focus on high-$z$ quasars at $1.82 \lesssim z \lesssim 2.16$ and with $\log L_{2200} > 45$, i.e., the same redshift range as \citet{Trammell2007} but 0.5 dex fainter at 2200~\angstrom~and a much larger sample of quasars with \galex detections. The top-left panel of Figure~\ref{fig:SED_bin} illustrates the rest-frame NUV to EUV median-normalized SEDs for our 3871 UV-detected quasars with $\log L_{2200} > 45$ up to 47.3 and separated into four luminosity bins. 
These SEDs, with the aforementioned observational biases uncorrected, do appear to be luminosity-dependent, consistent with \cite{Trammell2007} and \cite{Krawczyk2013}, but at a higher confidence level.

In Section~\ref{sect:perform_corrections}, we introduce corrections against two important observational biases, including the mis-matched bias and UV detection incompleteness bias. As we illustrate in Figure~\ref{fig:SED_bin}, the second bias has a much more pronounced effect. We present and discuss below results after correcting the biases with two strategies (see Sections~\ref{sect:uniform_sed} and \ref{sect:different_matching_strategy}), indicating the mean UV SED is likely universal (luminosity-independent) beyond $\log L_{2200} > 45$. Implication of the universal mean UV SED on the Baldwin effect is discussed in Sections~\ref{sect:baldwin_effect}.
Furthermore, in Section~\ref{sect:intrinsic_mean_SED}, we then obtain an intrinsic mean UV SED of a sub-sample of quasars with high UV detection (most luminous and with deepest \galex exposure) for which an absolute correction to the UV detection incompleteness bias is possible by taking the \galex detection limits into account, and compare the intrinsic bias-free mean UV SED with previous spectroscopic results in Section~\ref{sect:previous_results}. Finally, implication on the accretion flow is discussed in Section~\ref{sect:physical_implication}.

\subsection{A universal mean UV SED at $\log L_{2200} > 45$ up to 47.3}\label{sect:uniform_sed}

\begin{figure}[t!]
\centering
\includegraphics[width=0.45\textwidth]{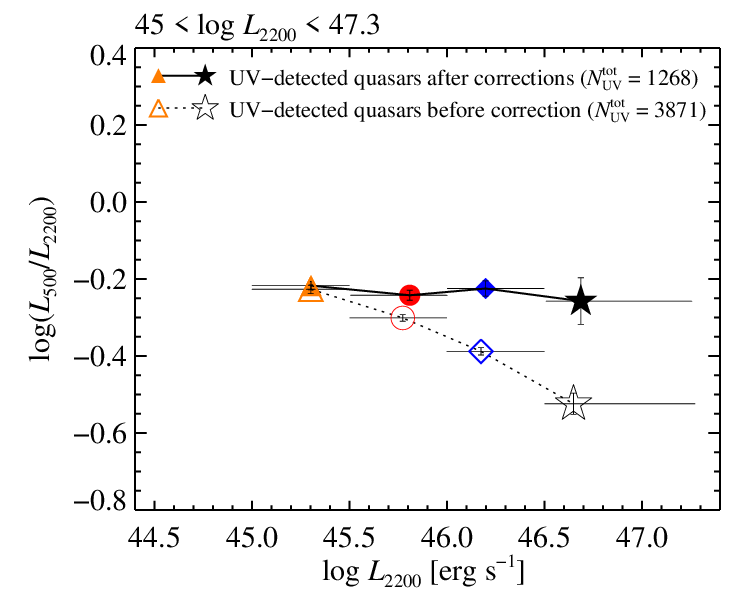}
\caption{
The extent of the EUV hardness, indicated by $\log(L_{500}/L_{2200})$, versus the 2200 \angstrom~luminosity for the UV-detected quasar sample with $45 < \log L_{2200} < 47.3$, before (open symbols linked by dotted line) and after (filled symbols linked by solid line) bias corrections. The vertical error bars indicate the same 1$\sigma$ statistical error as Figure~\ref{fig:SED_bin}, while the horizontal bars indicate the luminosity range of quasars in each luminosity bin.
}\label{fig:correction_extent_on_luminosity}
\end{figure}

\begin{figure*}[t!]
\centering
\includegraphics[width=0.35\textwidth]{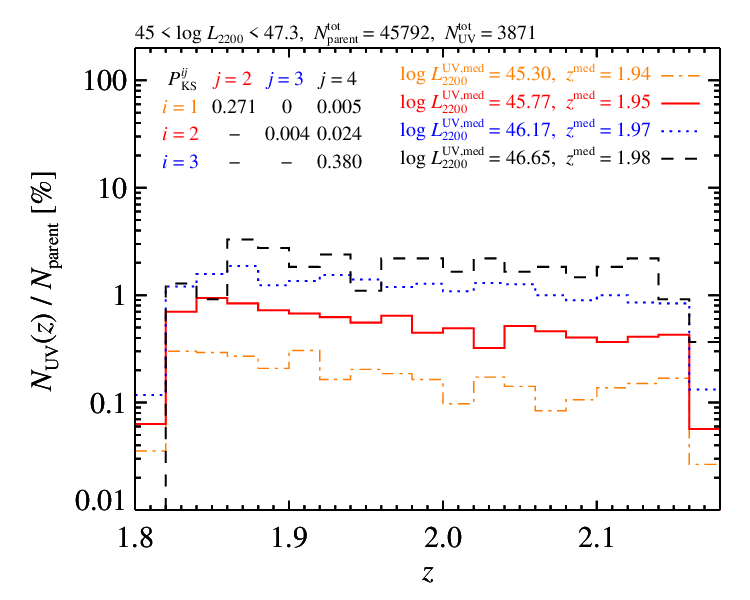}
\includegraphics[width=0.35\textwidth]{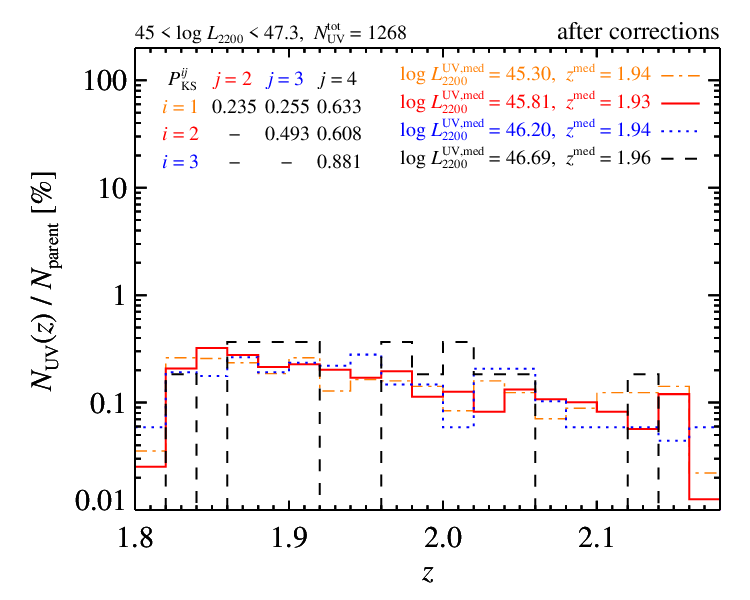}
\includegraphics[width=0.35\textwidth]{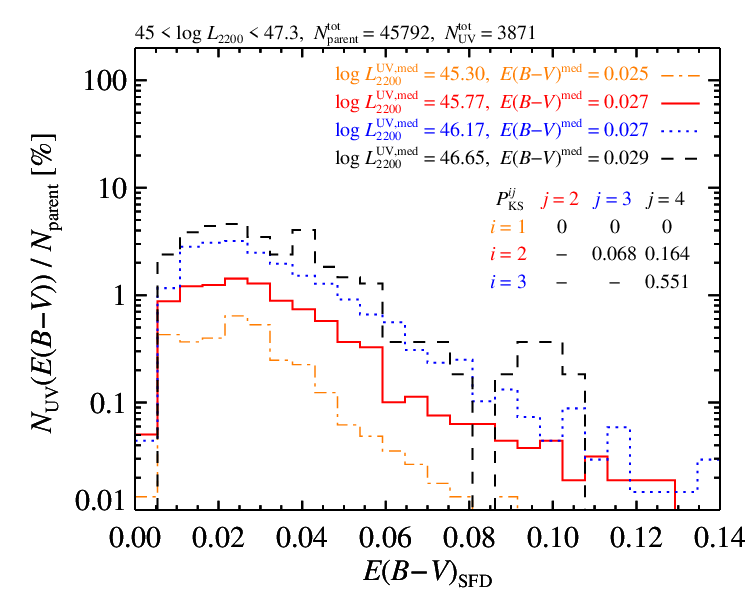}
\includegraphics[width=0.35\textwidth]{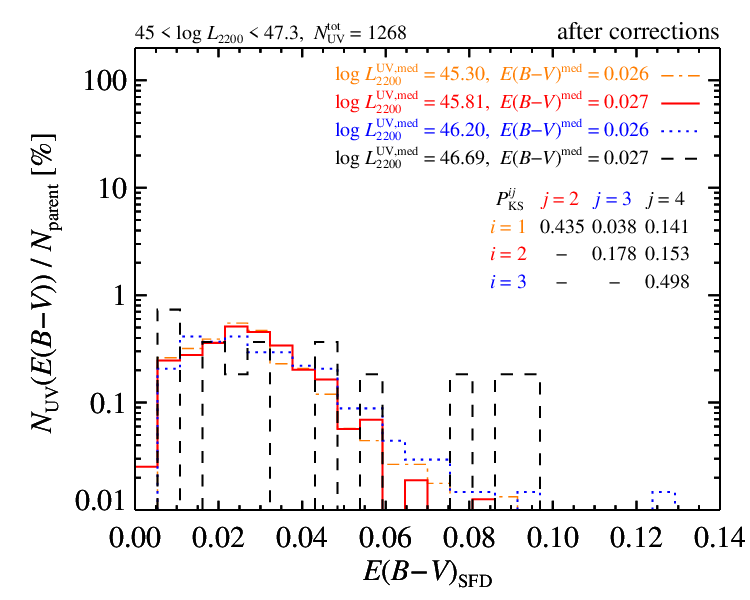}
\includegraphics[width=0.35\textwidth]{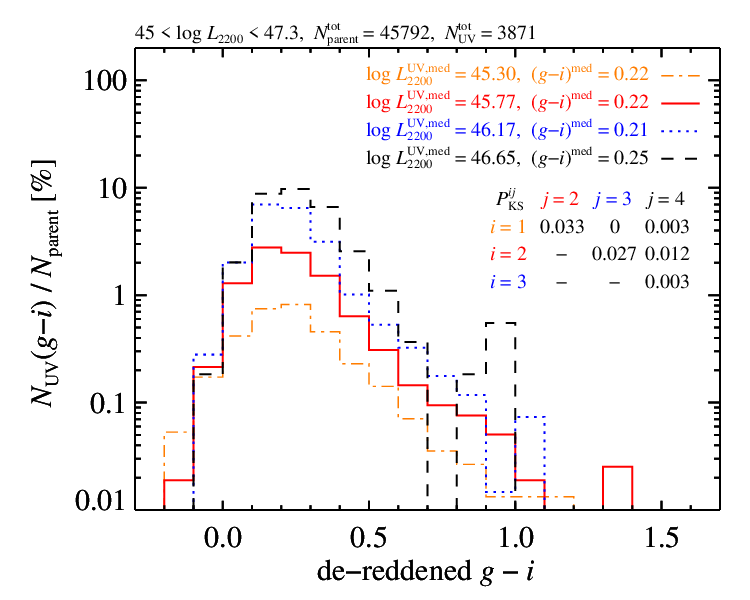}
\includegraphics[width=0.35\textwidth]{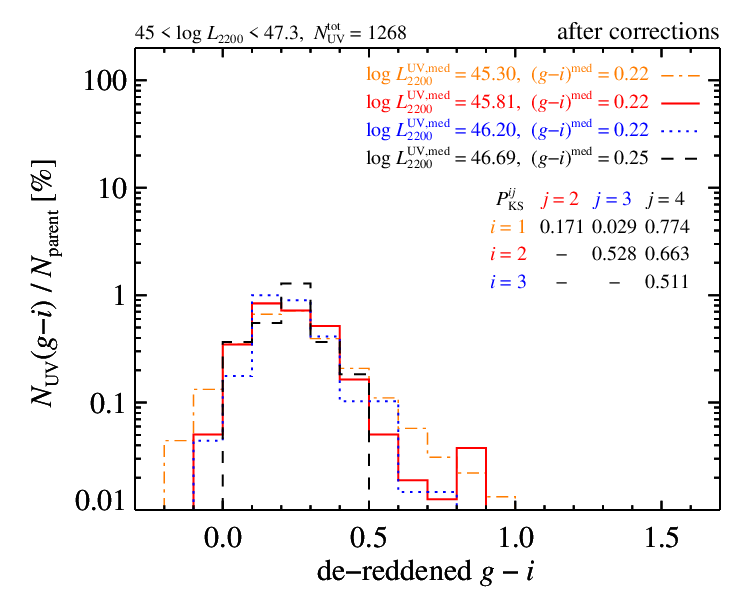}
\includegraphics[width=0.35\textwidth]{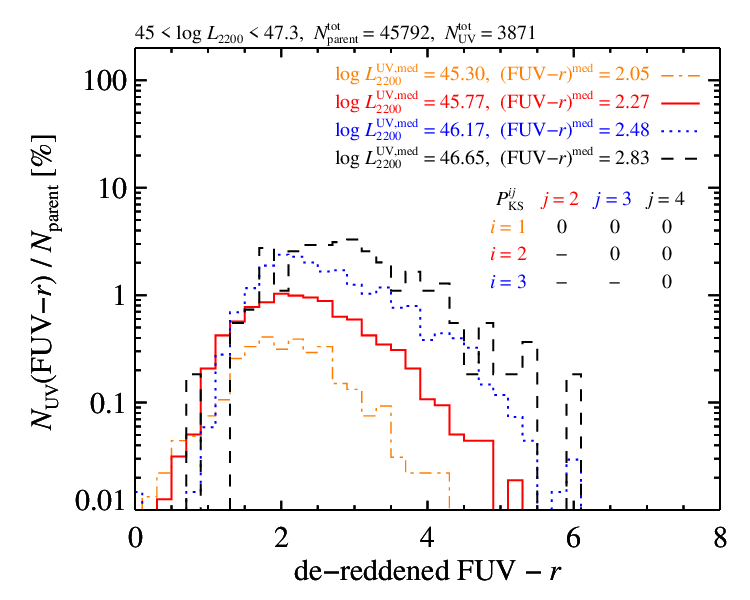}
\includegraphics[width=0.35\textwidth]{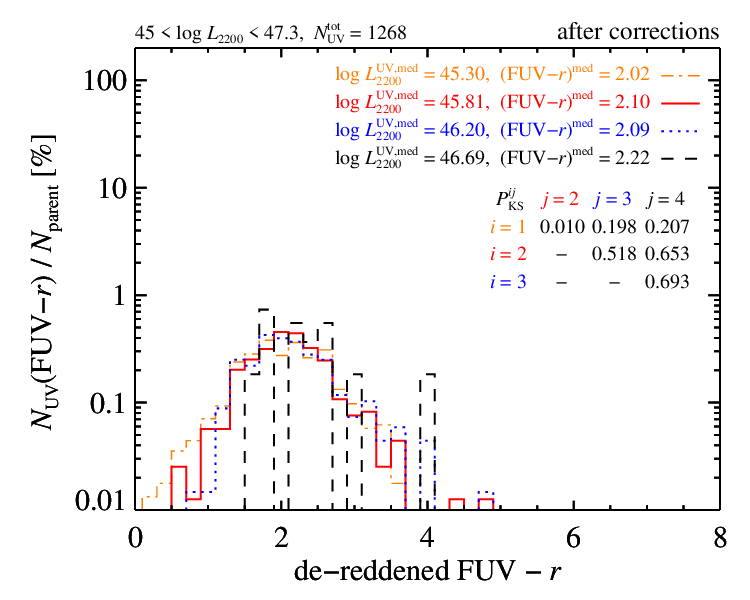}
\caption{
Relative distributions of redshift (top-row panels), $E(B-V)_{\rm SFD}$ (second-row panels), de-reddened $g-i$ color (third-row panels), and de-reddened FUV-$r$ color (bottom-row panels) for quasar sample at $1.82 \lesssim z \lesssim 2.16$ and with $45 < \log L_{2200} < 47.3$, before (left column) and after (right column) bias corrections. Here four luminosity bins are considered: $45 < \log L_{2200} < 45.5$ (orange dot-dashed lines), $45.5 < \log L_{2200} < 46$ (red solid lines), $46 < \log L_{2200} < 46.5$ (blue dotted lines), and $46.5 < \log L_{2200} < 47.3$ (black dashed lines). In each panel, legends contain median values of the 2200~\angstrom~luminosity and the corresponding quantities. The probabilities of the K-S test, $P^{ij}_{\rm KS}$, indicate the differences among any two distributions, where $i~({\rm or}~j) = 1$ stands for the adopted faintest luminosity bin and increases for brighter luminosity bin. For clarify, $P^{ij}_{\rm KS} < 0.001$ are simply nominated as zero.
}\label{fig:dis_z_EBV_color_before_after_corrections}
\end{figure*}

\begin{figure*}[t!]
\centering
\includegraphics[width=0.9\textwidth]{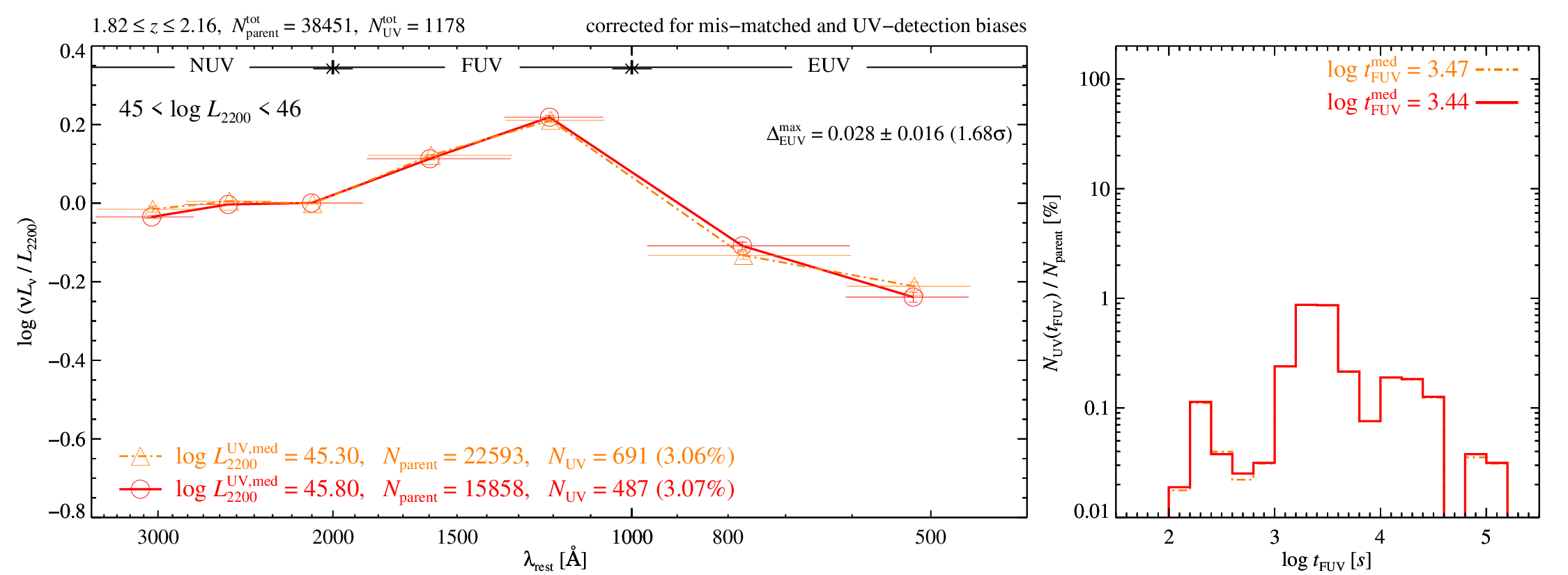}
\includegraphics[width=0.9\textwidth]{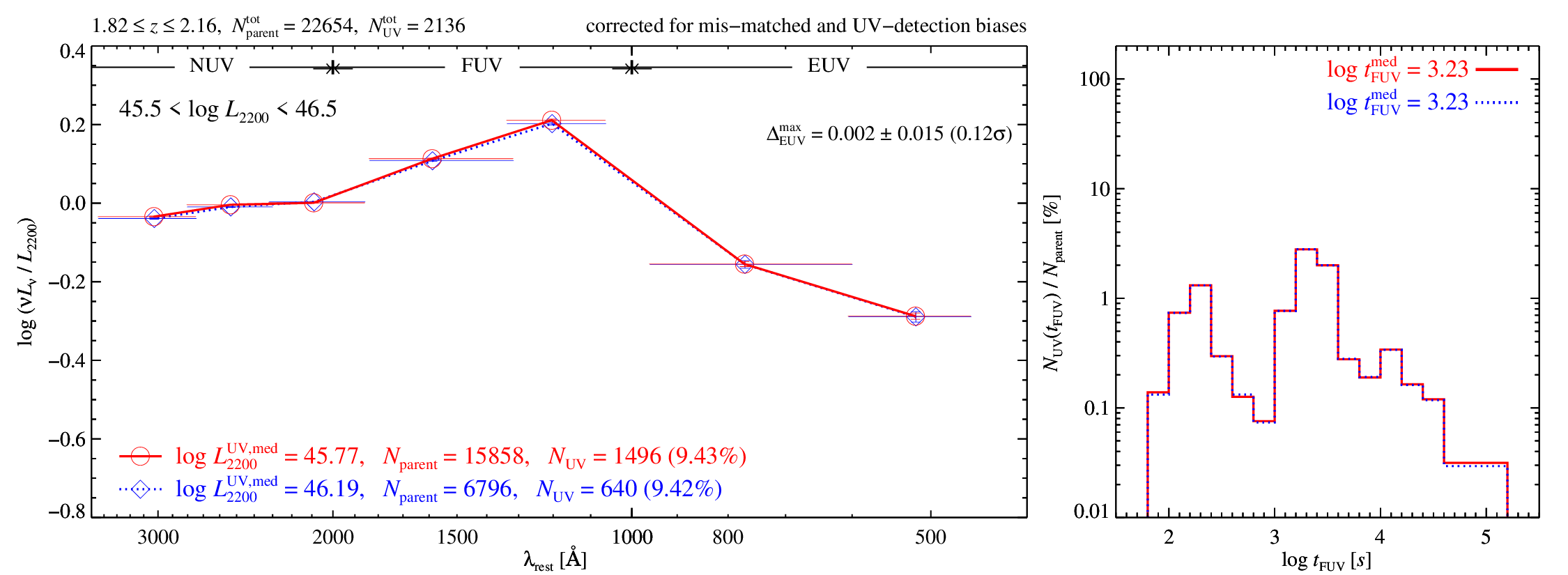}
\includegraphics[width=0.9\textwidth]{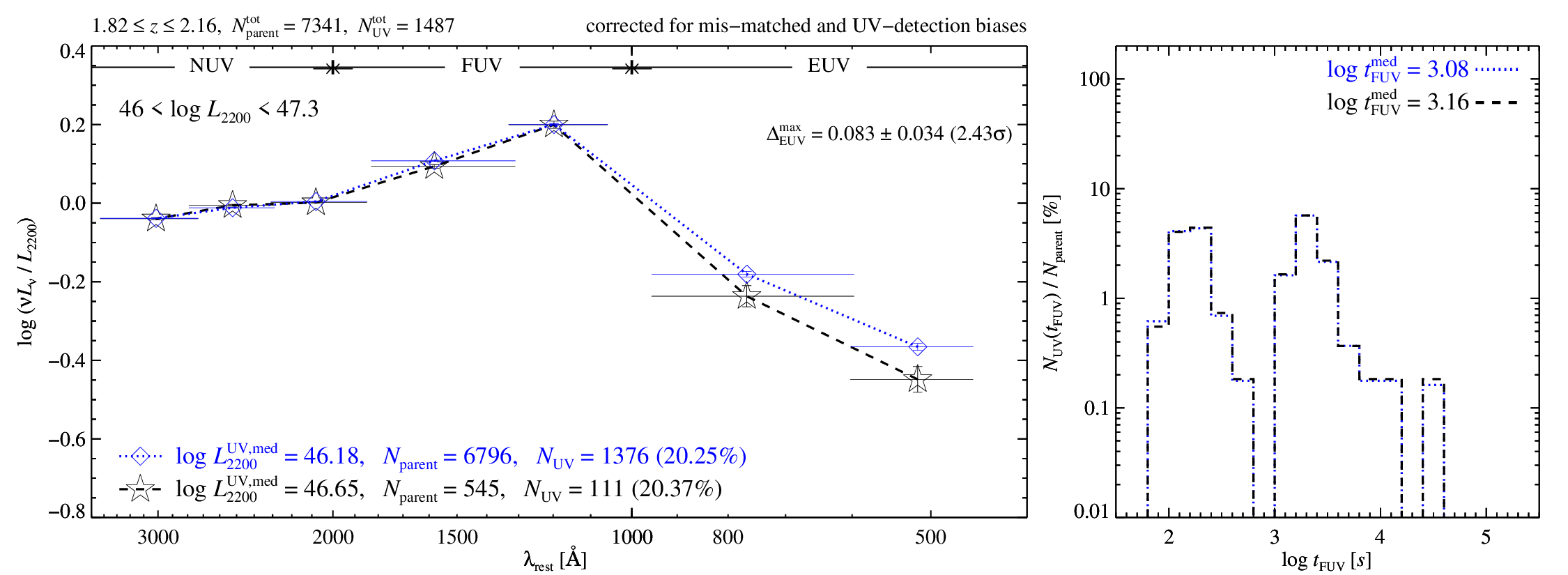}
\caption{
Panels in each row are the same as the bottom-row panels of Figure~\ref{fig:SED_bin}, but for quasar samples within adjacent two luminosity bins: $45 < \log L_{2200} < 46$ (top row), $45.5 < \log L_{2200} < 46.5$ (middle row), and $46 < \log L_{2200} < 47.3$ (bottom row).
}\label{fig:SED_bin_correct_different_limits}
\end{figure*}

After correcting against both the mis-matched and UV detection incompleteness biases, we interestingly find that the median-normalized SEDs down to the rest-frame $\sim 500$ \angstrom~are nearly luminosity-independent for $45 < \log L_{2200} < 47.3$ (see the bottom-left panel of Figure~\ref{fig:SED_bin}).
In Figure~\ref{fig:correction_extent_on_luminosity} we further illustrate the dependence of EUV SED slope (quantified with $\log(L_{500}/L_{2200})$) on luminosity, before and after our corrections. 

Observationally, we expect quasars with higher $L_{2200}$ are more likely detected in the rest-frame EUV. In other words, given an EUV detection limit, quasars with higher $L_{2200}$ could be detected in the EUV with larger redshift, stronger Galactic extinction, or redder colors. 
Indeed, as shown in the left panels of Figure~\ref{fig:dis_z_EBV_color_before_after_corrections}, this is confirmed by the slightly increasing median redshift, $E(B-V)_{\rm SFD}$, and de-reddened $g-r$ and FUV-$r$ colors for the initial UV-detected quasar sample with increasing luminosity. Among different luminosity bins, the Kolmogorov-Smirnov test (hereafter, shortly the KS test) essentially indicates significant differences among the corresponding distributions. 
Impressively, as shown in the right panels of Figure~\ref{fig:dis_z_EBV_color_before_after_corrections}, the differences in these distributions disappear after bias corrections. This is remarkable as when applying the UV detection bias correction, we only match the UV detection fraction between luminosity bins and simultaneously require that quasar samples among different luminosity bins have comparable distribution of the \galex FUV exposure time (see the bottom-right panel of Figure~\ref{fig:SED_bin}), indicating the success of the bias correction.

The observed EUV SED of quasars could be affected by intergalactic absorption and intrinsic extinction.
For our high-redshift quasars at $z \sim 2$, the effects of intergalactic Lyman absorption of intervening systems, such as the \Lya~forest, Lyman limit systems, and damped \Lya~absorbers, could be significant \citep[e.g.,][]{Prochaska2014,Lusso2015}. 
The narrow ($z \sim 2 \pm 0.2$) and similar redshift distributions of our bias-corrected quasar samples in different luminosity bins (see the top-right panel of Figure~\ref{fig:dis_z_EBV_color_before_after_corrections}) however suggest that further applying a correction for the intergalactic absorption (even if possible for individual quasars) would not change our conclusion on the luminosity-independent mean UV SED.
The third- and bottom-right panels of Figure~\ref{fig:dis_z_EBV_color_before_after_corrections} show the $g-i$ (i.e., the rest-frame NUV to FUV) and FUV-$r$ (i.e., the rest-frame NUV/FUV to EUV) colors for our bias-corrected quasar samples, respectively. The indistinguishable distributions of these colors among different luminosity bins suggest the intrinsic extinctions in these bins are also globally comparable.

The intrinsic variation of quasars may also affect the observed SED, and non-simultaneous observations could yield additional scatter to the observed SED slope distribution. 
Quasars generally have larger variability amplitudes at shorter UV wavelengths \citep[e.g.,][]{Welsh2011,Zhu2016}, and the EUV variation amplitude appears luminosity-dependent, i.e.,  decreasing with increasing luminosity \citep[][]{Welsh2011}.
Then, the variability could yield a biased SED slope bluer than intrinsic, as quasars in their EUV brighter status by coincidence are more likely detected by {\it GALEX}. Correcting this bias is non-straightforward as the EUV variation measurements are unavailable for most of our quasars. However, the effect of variability is expected to be subtle.
First, the intrinsic variability only partly contributes to the scatter of EUV SED. Second, the variability may only show weak dependence on luminosity \citep[e.g.,][]{Ai2010} and our sample spans a limited range of luminosity. If ignoring the luminosity-dependence of the variability, the variability would yield comparable bias in various luminosity bins, and the bias would be mostly canceled out when comparing the SED between luminosity bins. The residual bias due to luminosity-dependence of the variability amplitude is thus expected to be weak. 
Last, correcting such weak residual bias could yield redder EUV SED in lower luminosity bin (where the variation and thus the bias caused by variation are stronger) than in higher luminosity bin, contrary to the directly observed trend (see the top-left panel in Figure~\ref{fig:SED_bin}). 

In sum, the observed EUV slopes simply derived for quasars with \galex FUV detections are redder for more luminous quasars, which is due to prominent observational bias, primarily the UV detection incompleteness bias. After correcting against the biases, we find that at $\log L_{2200} > 45$ the mean quasar SEDs are nearly luminosity-independent. Such conclusion is robust against observational effects including intergalactic absorption, intrinsic extinction, and intrinsic variation.

\subsection{A different matching strategy}\label{sect:different_matching_strategy}

We note that to correct the UV detection incompleteness bias, we need to drop many UV detections in higher luminosity bins with higher \galex detection fraction. 
To maximize the number statistics, in this subsection (Section~\ref{sect:different_matching_strategy}), instead of matching all bins simultaneously with a fixed lowest luminosity bin, we adopt a different matching strategy through matching two adjacent luminosity bins, between which the difference in the UV detection fractions would not be too large.

The results are presented in Figure~\ref{fig:SED_bin_correct_different_limits}.
As expected, much more sources are kept in the high luminosity bins and then at higher significance level we confirm the luminosity-independent mean quasar SED, except in the brightest two luminosity bins (i.e., the bottom-left panel of Figure~\ref{fig:SED_bin_correct_different_limits} for $46 < \log L_{2200} < 47.3$). For the later case, we note that the EUV difference if any is less than 0.1 dex and being limited by the quite few quasar numbers in the brightest luminosity bin the significance of the difference is less than $3\sigma$. Although a larger quasar sample would be vital to settle up this, we speculate the universal mean quasar SED may be there for all quasars brighter than $\log L_{2200} \simeq 45$.

The universal mean quasar SED discussed up to now is only argued for quasars brighter than $\log L_{2200} \simeq 45$. Resolving whether or not this universality can persist, even to the fainter luminosity regime, undoubtedly requires a more complete and fainter UV-detected quasar sample from future surveys. For instance, the China Space Station Telescope (CSST) would perform both wide and deep field surveys, reaching average 5$\sigma$ NUV depths of AB $\simeq 25 - 26$ mag \citep{Zhan2018,Zhan2021}, and more excitingly, the {\it Ultraviolet Explorer} (UVEX) will probe the time domain NUV and FUV all-sky surveys, reaching both 5$\sigma$ NUV and FUV depths of AB $\simeq 25$ mag, $\geqslant 50$ times deeper than \galex \citep{Kulkarni2021}.

\subsection{The universal mean UV SED versus the Baldwin Effect}\label{sect:baldwin_effect}

\begin{figure*}[t!]
\centering
\includegraphics[width=0.4\textwidth]{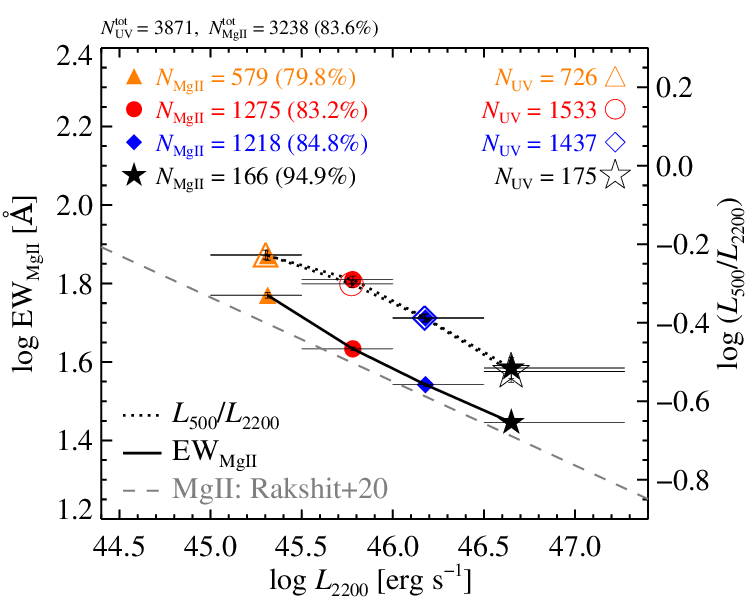}
\includegraphics[width=0.4\textwidth]{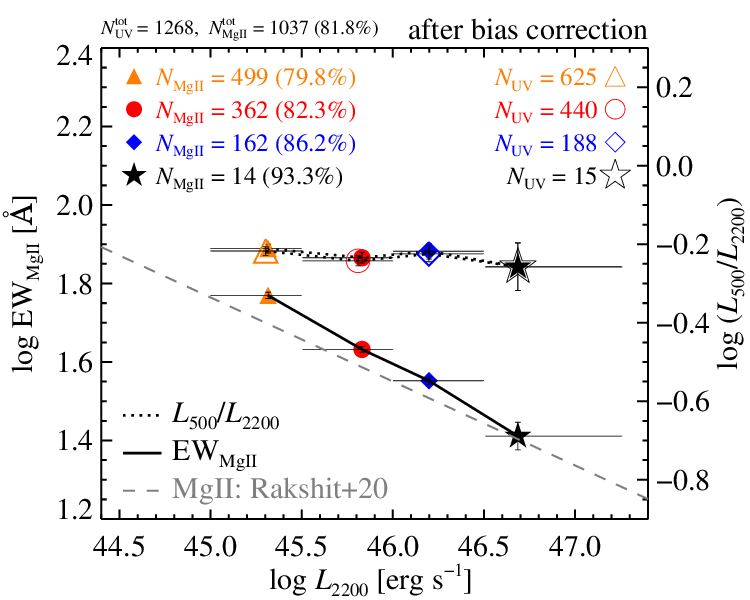}
\includegraphics[width=0.4\textwidth]{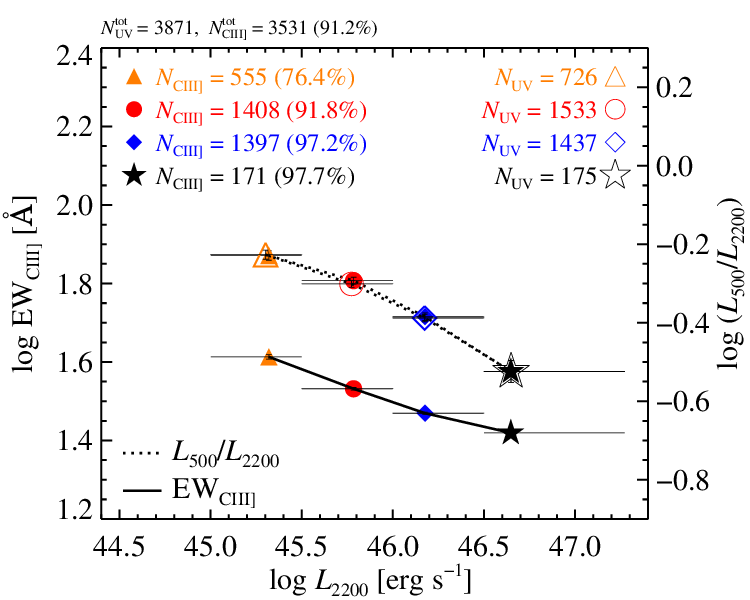}
\includegraphics[width=0.4\textwidth]{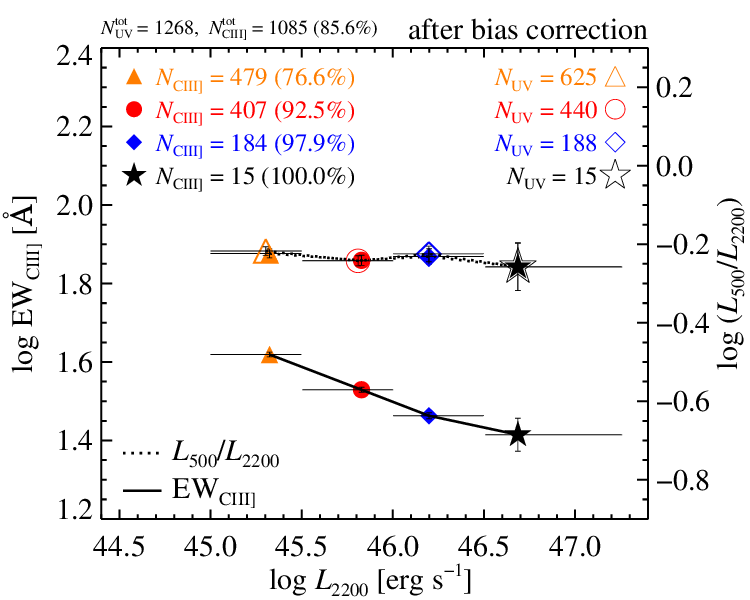}
\includegraphics[width=0.4\textwidth]{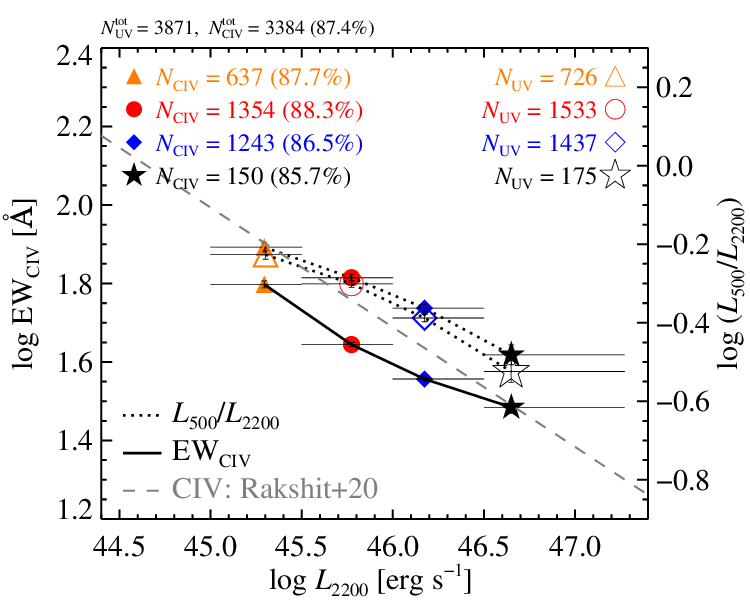}
\includegraphics[width=0.4\textwidth]{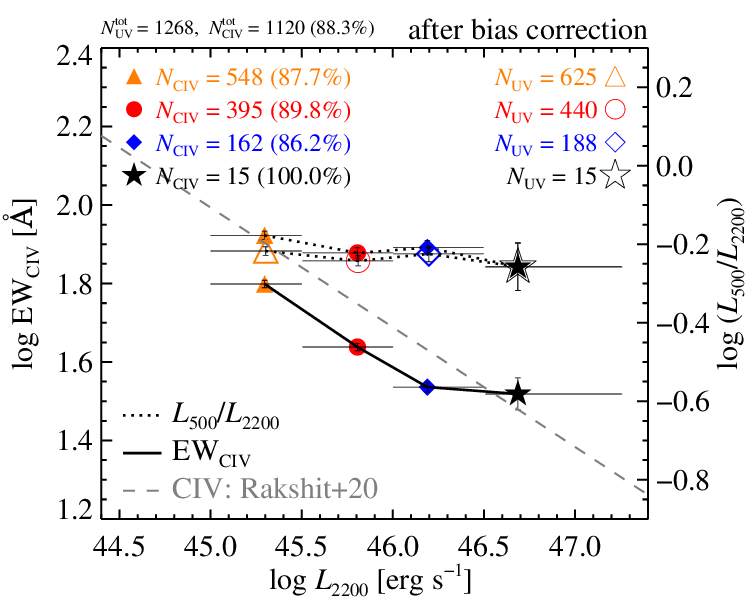}
\caption{
Top panels: The \MgII~EW (solid lines; left axis) and $\log(L_{500}/L_{2200})$ (dotted lines; right axis) versus the 2200~\angstrom~luminosity for quasars with $\log L_{2200} > 45$. In each luminosity bin, the open symbol presents the median of $\log(L_{500}/L_{2200})$ for the UV-detected quasars with $N_{\rm UV}$ sources, while the filled symbols present the medians of $\log(L_{500}/L_{2200})$ and \MgII~EW for the UV-detected quasars with reliable measurement of the \MgII~line, nominated as $N_{\rm MgII}$ sources followed by a \MgII~fraction of $N_{\rm MgII}/N_{\rm UV}$ in the bracket. The 1$\sigma$ uncertainties (vertical error bars) are estimated as the standard deviation divided by the square root of quasar number in each luminosity bin, while the luminosity range of quasars in that bin is presented by the horizontal bar.
The left and right panels show the results before and after bias correction, respectively.
The correlation between the \MgII~EW and $L_{3000}$ of the DR14Q quasars fit by \citet{Rakshit2020} is shown for comparison after applying a global conversion of $\log L_{2200} = \log L_{3000} + 0.03$ assuming the mean quasar SED of \citet{Richards2006}.
Middle and bottom panels: same as the top ones, but for the \CIII~line and \CIV~line, respectively. 
}\label{fig:EW_MgII_L2200_logL45}
\end{figure*}

If the mean SED of quasars brighter than $\log L_{2200} \simeq 45$ are indeed identical (luminosity-independent), it would certainly have strong implications on the well-known Baldwin effect \citep{Baldwin1977}.
The Baldwin effect, claiming an anti-correlation between the equivalent width (EW) of emission line and the continuum luminosity, has been well established for many emission lines of AGN \citep[e.g.,][]{Kinney1990,OsmerShields1999,Green2001,Dietrich2002,Dong2009,Rakshit2020}. However, its physical origin is still unclear. Several physical properties have been proposed to be luminosity dependent to account for this effect, including the covering factor of optically thin line-emitting clouds \citep{Shields1995}, the ionization parameter as well as covering factor of clouds \citep{MushotzkyFerland1984}, the chemical composition of gas \citep{Korista1998}, and the shape of ionizing EUV--soft X-ray continuum \citep{Netzer1992,ZhengMalkan1993}. 
Among these explanations, the softening of EUV--soft X-ray continuum with increasing luminosity is the most attractive one. For example, \citet{Trammell2007} claim that the intrinsic differences in the strength of the ionizing continuum implied by the observed luminosity dependence of the EUV SEDs may account for the Baldwin effect. As we have demonstrated in previous sections, the differences of the EUV SEDs are probably biased by selection effects and can fade away after bias correction (Sections~\ref{sect:uniform_sed} and \ref{sect:different_matching_strategy}). Therefore, it is interesting to further examine the role of the ionizing EUV continuum on accounting for the Baldwin effect, using emission lines available in the \citet{Rakshit2020} catalog.

In our concerned redshift range, the emission lines available in the \citet{Rakshit2020} catalog include \MgIILambda, \CIIILambda, \CIVLambda, and \LyaLambda, and considering our final parent 45,792 quasars (Section~\ref{sect:sample_summary}), the fractions of sources with reliable line measurements, defined by combining their quality flag $= 0$ and median $S/N$ per pixel $\geqslant 1$, are $\simeq 76.6\%$, $\simeq 78.1\%$, $\simeq 81.7\%$, and $\simeq 4.6\%$ for these lines, respectively. 
To explore the relationship between the Baldwin effect and the ionizing EUV continuum, we need to focus on the UV-detected quasar sample. For the UV-detected 3871 quasars with $\log L_{2200} > 45$ (Section~\ref{sect:sample_summary}), the fractions of sources with reliable \MgII, \CIII, \CIV, and \Lya~line measurements are 83.6\%, 91.2\%, 87.4\%, and 3.5\%, respectively. We confirm that the Baldwin effects of these lines implied by both the parent and the UV-detected quasar samples are almost the same.
Since the fraction of reliable \Lya~line is quite low, we would not further consider it in the following. 

Note for the collisionally excited \CIV~line, its ionization energy is as high as 47.9 eV, or less than 259~\angstrom, which is slightly shorter than the minimal wavelength covered by our sample. 
For the collisionally excited \CIII~line, its ionization energy is 24.4 eV (or 508~\angstrom), which is just around the minimal wavelength covered by our sample. 
For the \MgII~line, if it is mostly collisionally excited \citep[e.g.,][]{Dietrich2002,Guo2020}, it has quite low ionization energy, i.e., 7.6 eV (or 1631~\angstrom); instead, if it has a recombination origin, its ionization energy is 15.0 eV (or 827~\angstrom). Therefore, the ionizing continuum responsible for the \MgII~line is well covered by our sample and so it is very suitable for discussing the implication of the luminosity-independent mean UV SED on the Baldwin effect. 

In Figure~\ref{fig:EW_MgII_L2200_logL45}, we quantitatively compare the luminosity dependence of the \MgII~ (top panel), \CIII~(middle), and \CIV~(bottom) EWs with the luminosity dependence of the mean EUV SEDs, for the UV-detected quasars with $\log L_{2200} > 45$. Here the mean EUV SED is characterized by $L_{500}/L_{2200}$.
Since not all quasars have reliable line measurements provided by \citet{Rakshit2020}, we first check the mean EUV SEDs of all UV-detected quasars and of those with available line measurements (see the filled and open symbols linked by the dotted lines in Figure~\ref{fig:EW_MgII_L2200_logL45}), but find negligible difference between them. This indicates that quasars with reliable line measurements are not biased to either bluer or redder EUV SEDs. 
Then, for those quasars with reliable line measurements, we examine their Baldwin effects of the \MgII~and \CIII~lines, finding that they have similar luminosity dependence to that of the mean EUV SEDs before our bias corrections (top- and middle-left panels in Figure~\ref{fig:EW_MgII_L2200_logL45}). As proposed by many previous studies \citep[e.g.,][]{Netzer1992,ZhengMalkan1993,Trammell2007}, the similar anti-luminosity dependence of both the line EWs and the mean EUV SEDs suggests that the softening of the ionizing continuum with increasing luminosity could be responsible for the Baldwin effect of the \MgII~(and \CIII) line. However, this statement is based on a quasar sample subject to selection biases.

We then repeat the comparison above but now for quasar samples after our bias corrections (as described in Section~\ref{sect:uniform_sed}).
As shown in the right panels of Figure~\ref{fig:EW_MgII_L2200_logL45}, the mean EUV SEDs, after bias corrections, are nearly independent to luminosity. However, the Baldwin effect of the lines preserves on both slope and amplitude before and after the correction. Note that in each luminosity bin the fractions of quasars with reliable line measurements before and after bias correction are more or less comparable. Therefore, our results indicate that the shape of ionizing EUV continuum is not the main driver for the Baldwin effect, at least for the \MgII~and \CIII~lines, and probably also for \CIV\ line whose ionization potential is beyond the spectral coverage of EUV SEDs we obtained. Instead, the Baldwin effect should closely reflect some luminosity-dependent properties of the line-emitting clouds, such as the chemical composition of gas \citep{Korista1998} or the covering factor \citep[e.g.,][]{Dong2009}. 

Here we propose another possibility based on a recently unveiled correlation between the emission line properties and the amplitude of optical variation \citep{Kang2021, RenWK2022}. Briefly, more variable quasars have stronger emission lines, remaining robust after controlling the effects of bolometric luminosity, BH mass, Eddington ratio, and redshift. Since the UV/optical variation is likely driven by accretion disk turbulence \citep{DexterAgol2011,Cai2016,Cai2018,Cai2020b}, \citet{Kang2021} propose two potential underlying mechanisms: stronger disk turbulence could lead to either a bluer EUV SED \citep[cf. Figure 4 of][]{Cai2016} or the launch of more clouds with larger covering factors \citep{CzernyHryniewicz2011}, both of which can result in stronger emission lines. 
Considering the long-term optical variation amplitude SF$_{\infty}$ is generally found to be anti-correlated with luminosity, e.g., for the $i$-band absolute magnitude $M_i$ approximately $\log {\rm SF}_{\infty} \propto 0.1 M_i$ \citep{MacLeod2010,Guo2017}, our universal mean SEDs across $\sim 2$ dex in the UV luminosity then potentially prefer stronger disk turbulence launching more clouds as the main origin of the Baldwin effect. 
For more luminous AGN, their accretion disks are more stable, probably owing to stronger magnetic fields \citep[e.g.,][]{Cai2019}, and so less turbulence, resulting in fewer line-emitting clouds and smaller line EWs.
Last but not least is another observational fact that the Baldwin effect of higher ionization lines tends to be stronger \citep[e.g.,][]{Zheng1992,Dietrich2002}.
In the aforementioned turbulence scenario, this may implies that the launch of high-ionization line-emitting clouds at small radii could be more sensitive to the disc turbulence.
This new perspective analyzing the interrelationship to AGN variability provides a new clue on the origin of the Baldwin effect. However, more complete and fainter quasar samples are crucial to make a final conclusion. Remarkably, future time domain AGN surveys in optical, to be conducted by such as the northern-sky Wide Field Survey Telescope (WFST) and the southern-sky Large Synoptic Survey Telescope \citep[LSST;][]{LSSTScienceCollaboration2009-SB,LSSTScienceCollaboration2017-OS,Brandt2018-LSST-DDF}, would provide a great amount of measurements on multi-wavelength variation amplitudes of quasars in the observed-frame optical. 
Being complemented with the CSST NUV all-sky survey \citep{Zhan2018,Zhan2021} and the UVEX NUV/FUV time domain survey \citep{Kulkarni2021}, it would be valuable in exploring deeper AGN and accretion physics, such as those discussed above and in the following sub-sections.

\subsection{An intrinsic bias-free mean UV SED}\label{sect:intrinsic_mean_SED}

\begin{deluxetable}{lcrrr}
\tablecaption{The number of quasars brighter than three luminosity thresholds and covered by \galex tiles with three minimal FUV exposure times (see also Figure~\ref{fig:UV_detected_fractions_different_minimal_exposure_times_tFUV1_strict_longestFUV}). \label{tab:NUV_FUV_detection_limits}}
\tablewidth{0pt}
\tablehead{
\colhead{$t^{\rm min}_{\rm tile,FUV}/[s]$} & \colhead{$> \log L_{2200}$} & \colhead{$N_{> L}$} & \colhead{$N_{> L}^{\rm NUV}$} & \colhead{$N_{> L}^{\rm FUV}$}
}
\decimalcolnumbers
\startdata
  & 45.5 & 23199 & 14229 (61.3\%) & 4612 (19.9\%) \\
1 & 46.0 &  7341 &  5761 (78.5\%) & 2170 (29.6\%) \\
  & 46.5 &   545 &   485 (89.0\%) &  209 (38.3\%) \\
\hline
  & 45.5 & 6473 & 5003 (77.3\%) & 2206 (34.1\%) \\
{\bf 1000} & {\bf 46.0} & {\bf 2198} & {\bf 1897 (86.3\%)} &  {\bf 903 (41.1\%)} \\
  & 46.5 &  184 &  168 (91.3\%) &   74 (40.2\%) \\
\hline
     & 45.5 & 3035 & 2388 (78.7\%) & 1091 (35.9\%) \\
2000 & 46.0 &  906 &  792 (87.4\%) &  376 (41.5\%) \\
     & 46.5 &   76 &   69 (90.8\%) &   35 (46.1\%) \\
\enddata
\tablecomments{Tabulated in the third column are the numbers of quasars brighter than a given minimal UV luminosity (the second column) and covered by \galex tiles whose FUV exposure times are larger than a given minimal value (the first column). In the fourth and fifth columns, there are numbers of quasars with NUV or FUV detections (i.e., SNR$_{\rm NUV}>0$ or SNR$_{\rm FUV}>0$), respectively. Together, the numbers in parentheses are the corresponding NUV or FUV detection fractions defined as $N_{> L}^{\rm NUV}/N_{> L}$ or $N_{> L}^{\rm FUV}/N_{> L}$, respectively. The numbers in boldface are the reference values adopted to construct the bias-free mean EUV SED.}
\end{deluxetable}

\begin{figure}[t!]
\centering
\includegraphics[width=0.42\textwidth]{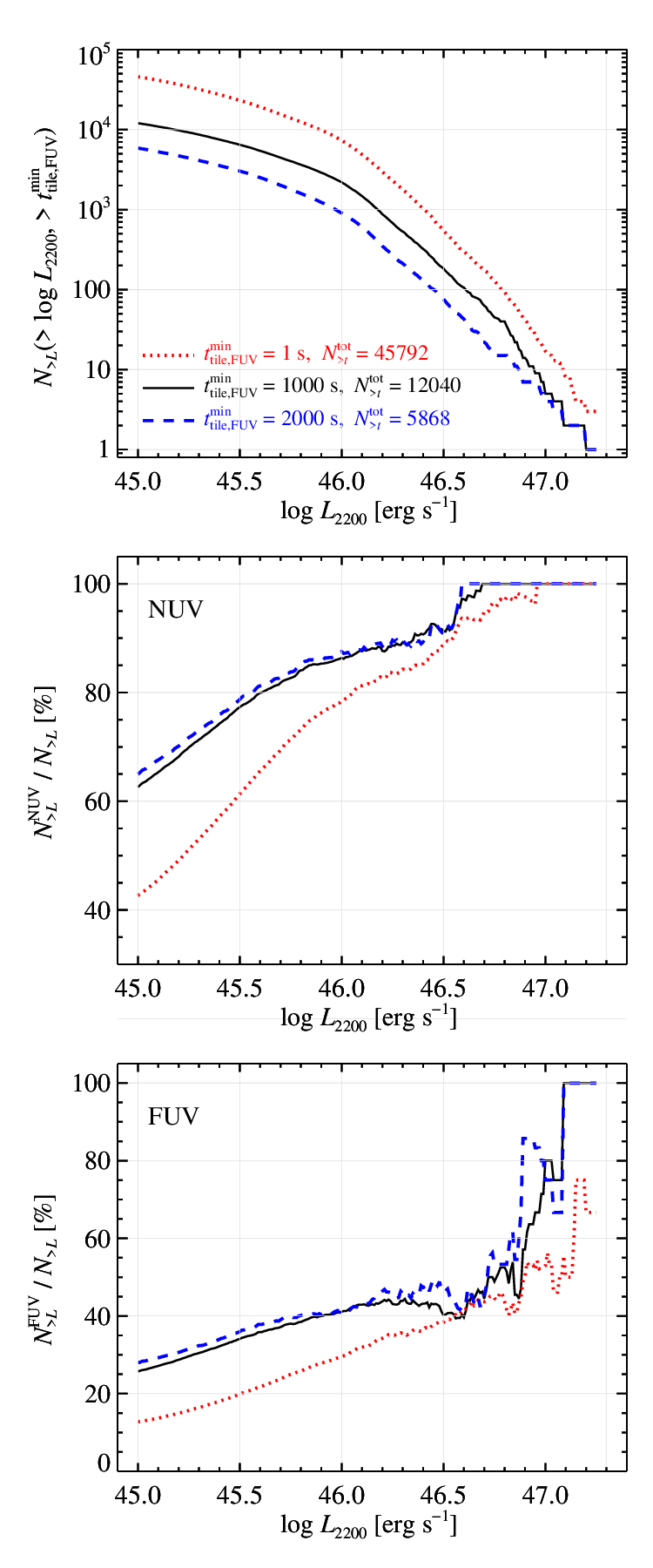}
\caption{
Top panel: being covered by \galex tiles with FUV exposure time larger than $t^{\rm min}_{\rm tile,FUV} = 1$ (red dotted line), 1000 (black solid line), and 2000 (blue dashed line) seconds, the numbers of quasars brighter than $\log L_{2200}$, $N_{> L}$, as a function of $\log L_{2200}$. The corresponding total numbers of quasars brighter than $\log L_{2200} = 45$, $N^{\rm tot}_{> t}$, are also nominated.
Middle panel: the resultant NUV detection fractions of quasars detected in the NUV band with SNR$_{\rm NUV} > 0$.
Bottom panel: same as the middle panel, but for the FUV detection fractions.
}\label{fig:UV_detected_fractions_different_minimal_exposure_times_tFUV1_strict_longestFUV}
\end{figure}

\begin{figure*}[t!]
\centering
\includegraphics[width=\textwidth]{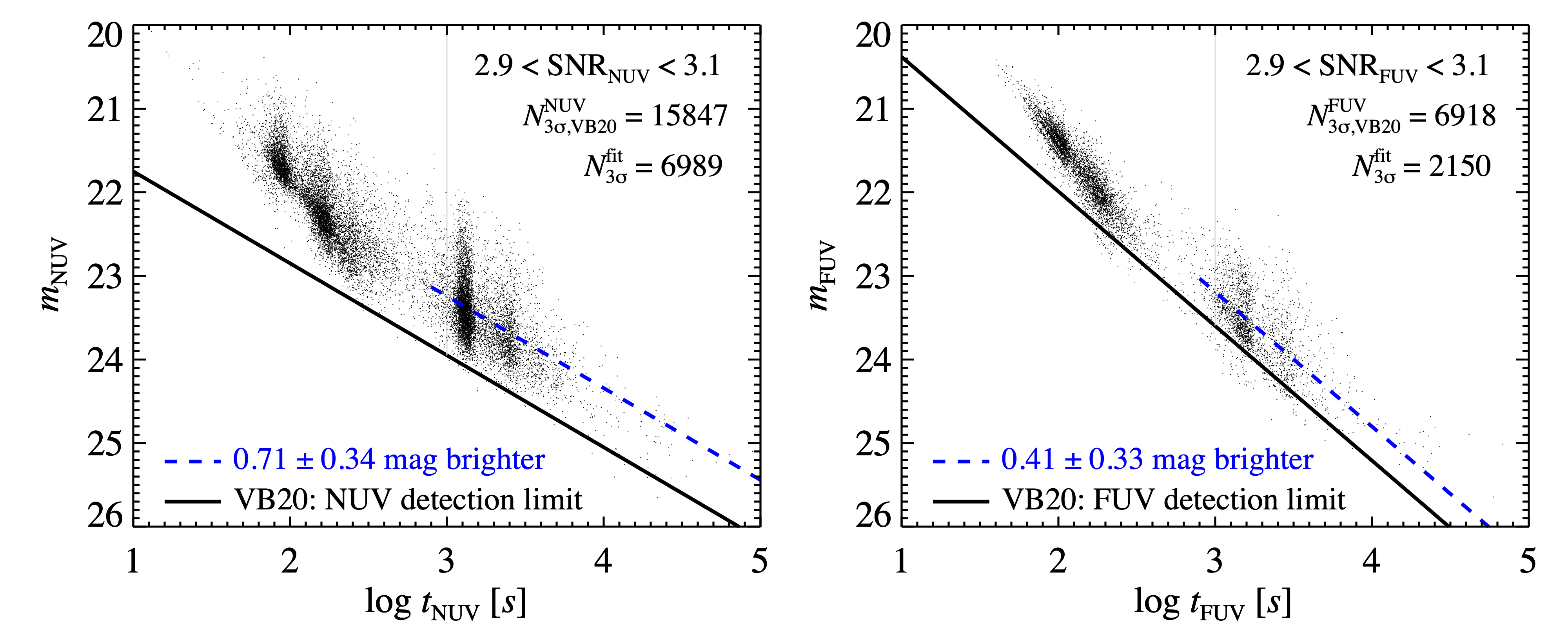}
\caption{The limiting magnitude of \galex images as a function of exposure time. 
Left panel: using the \citet[][VB20]{VandenBerk2020} catalog, tiny dots are quasars with \galex NUV detection at SNR$_{\rm NUV} = 3 \pm 0.1$ (the number of quasars labeled as $N^{\rm NUV}_{3\sigma,{\rm VB20}}$), compared to the 50\%-complete limiting magnitude given by \citet[][VB20; sold black line]{VandenBerk2020}. The $N^{\rm fit}_{3\sigma}$ quasars with NUV effective exposure times longer than 1000 seconds and SNR$_{\rm NUV} = 3 \pm 0.1$ are fit to derive the 3$\sigma$ NUV limiting magnitude as a function of exposure time (blue dashed line), assuming that the slope of the dependence of limiting magnitude on exposure time is the same as \citet{VandenBerk2020}.
The 3$\sigma$ limiting magnitude is brighter by 0.71 mag with 1$\sigma$ scatter of 0.34 mag than the 50\%-complete limiting magnitude of \citet{VandenBerk2020}. 
Right panel: same as the left panel, but for the \galex FUV band. Accordingly, the 3$\sigma$ FUV limiting magnitude is brighter by 0.41 mag with 1$\sigma$ scatter of 0.33 mag.
}\label{fig:detection_limits_NUV_FUV_texp_vb20}
\end{figure*}

\begin{figure*}[t!]
\centering
\includegraphics[width=\textwidth]{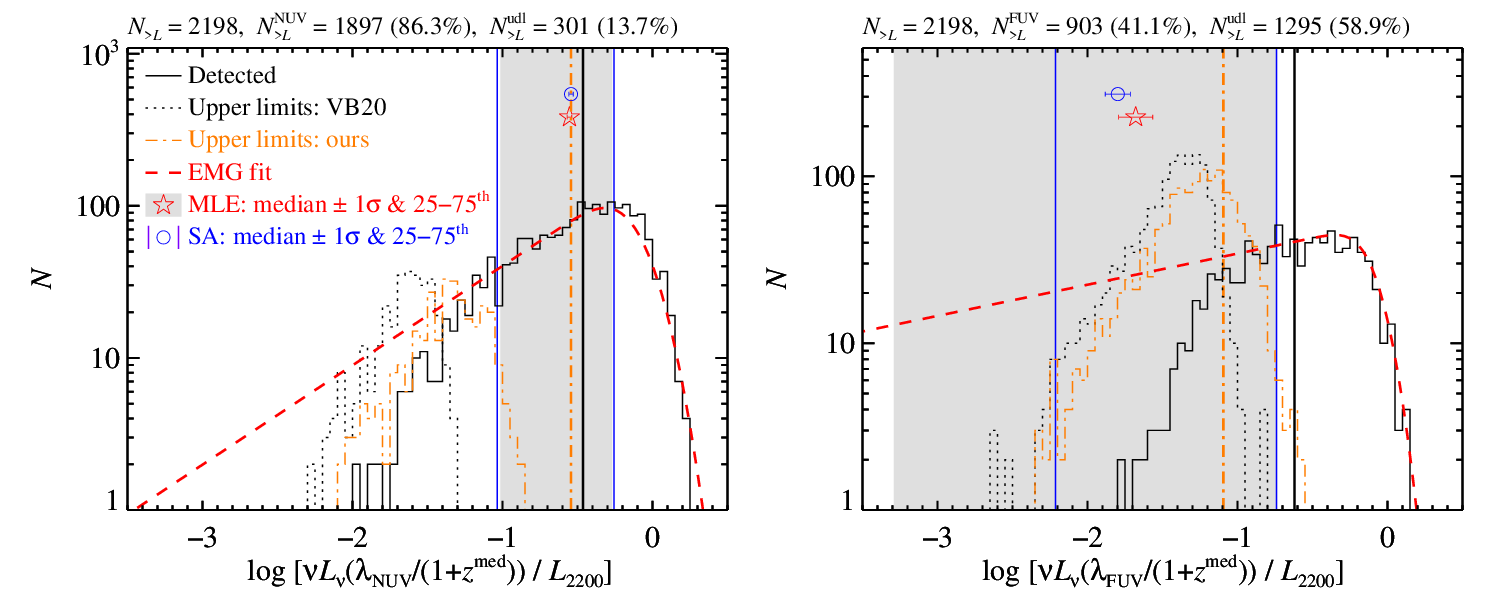}
\caption{
Left panel: Distributions of the rest-frame EUV luminosity (corresponding to the observed-frame NUV band for $z^{\rm med} \sim 2$) normalized to the rest-frame 2200~\angstrom, for $N_{>L}$ quasars brighter than $\log L_{2200} = 46$ and covered by \galex tiles with a minimal exposure time of 1000 seconds.
The black solid histogram is the distribution for the $N^{\rm NUV}_{>L}$ quasars with NUV detection (the black solid vertical line for the median). 
Shown for comparison are distributions for the $N^{\rm udl}_{>L}$ undetected quasars, assigned either the \citet[][VB20]{VandenBerk2020} upper detection limit (the red dotted histogram) or ours NUV upper detection limit (the orange dot-dashed histogram; the orange dot-dashed vertical line for the median of the whole bright sub-sample if our upper detection limits were the real detections).
Considering our upper detection limits, the median value (the blue open circle) and 25-75th percentile range (between the two blue vertical lines) inferred using the survival analysis \citep[SA;][]{FeigelsonNelson1985} are compared to those (the red open star and the light-gray region) inferred using the maximum-likelihood estimation (MLE) by fitting an exponentially modified Gaussian function \citep[EMG; the red dashed line for an intuitive fit;][]{VandenBerk2020}. 
The 1$\sigma$ uncertainties for the SA and MLE median values are estimated by bootstrapping 100 times the initial distribution (i.e., the black solid histogram plus the orange dot-dashed histogram).
Right panel: same as the left panel, but for the \galex FUV band.
}\label{fig:dis_detected_upperlimit_NUV_FUV}
\end{figure*}

\begin{figure*}[t!]
    \centering
    \includegraphics[width=0.49\textwidth]{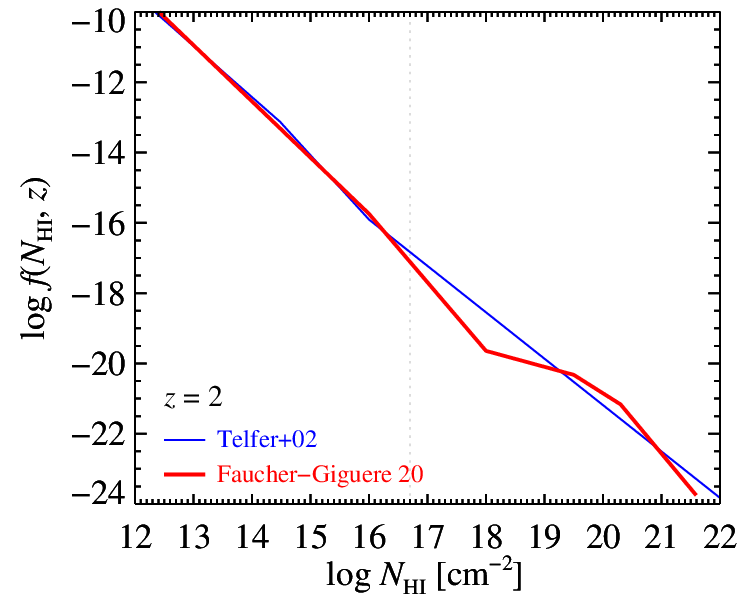}
    \includegraphics[width=0.49\textwidth]{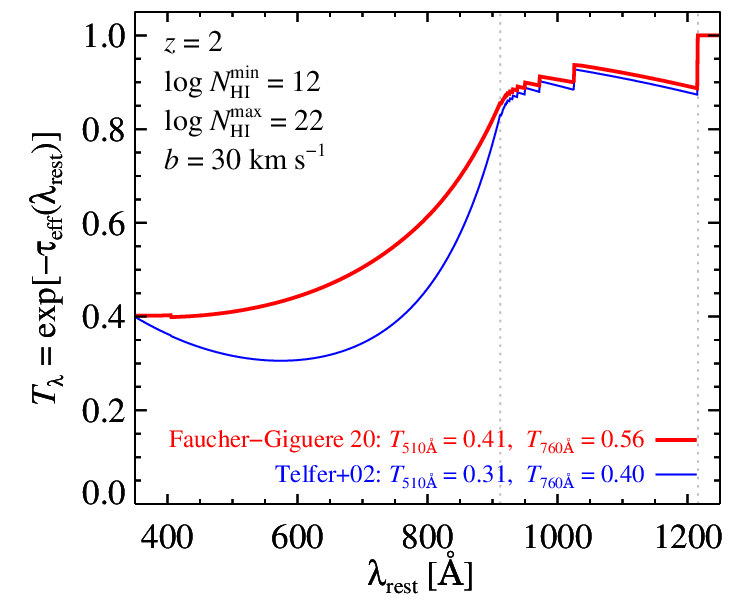}
    \caption{Left panel: distributions of absorbers at $z = 2$ for two parameterizations from \citet[][the thin blue line]{Telfer2002a} and \citet[][the thick red line]{Faucher-Giguere2020}. The light-gray vertical dotted line indicates a typical maximal $N_{\rm HI}$ adopted to correct the unresolved Lyman forest absorbers when constructing the IGM-corrected composite EUV spectrum \citep[e.g.,][]{Telfer2002a}.
    Right panel: the $z = 2$ IGM transmission curves as a function of rest-frame wavelength, adopting $b = 30~{\rm km~s^{-1}}$ and $12 \leqslant \log N_{\rm HI} \leqslant 22$, but distinct $f(N_{\rm HI}, z)$ from \citet[][the thin blue line]{Telfer2002a} and \citet[][the thick red line]{Faucher-Giguere2020}. At rest-frame 510~\angstrom~and 760~\angstrom, corresponding to the observed-frame \galex FUV and NUV bands for our quasars at $z \sim 2$, the IGM transmission values are nominated in the corresponding legends.}\label{fig:IGM_transmission}
\end{figure*}

\begin{figure*}[t!]
\centering
\includegraphics[width=0.8\textwidth]{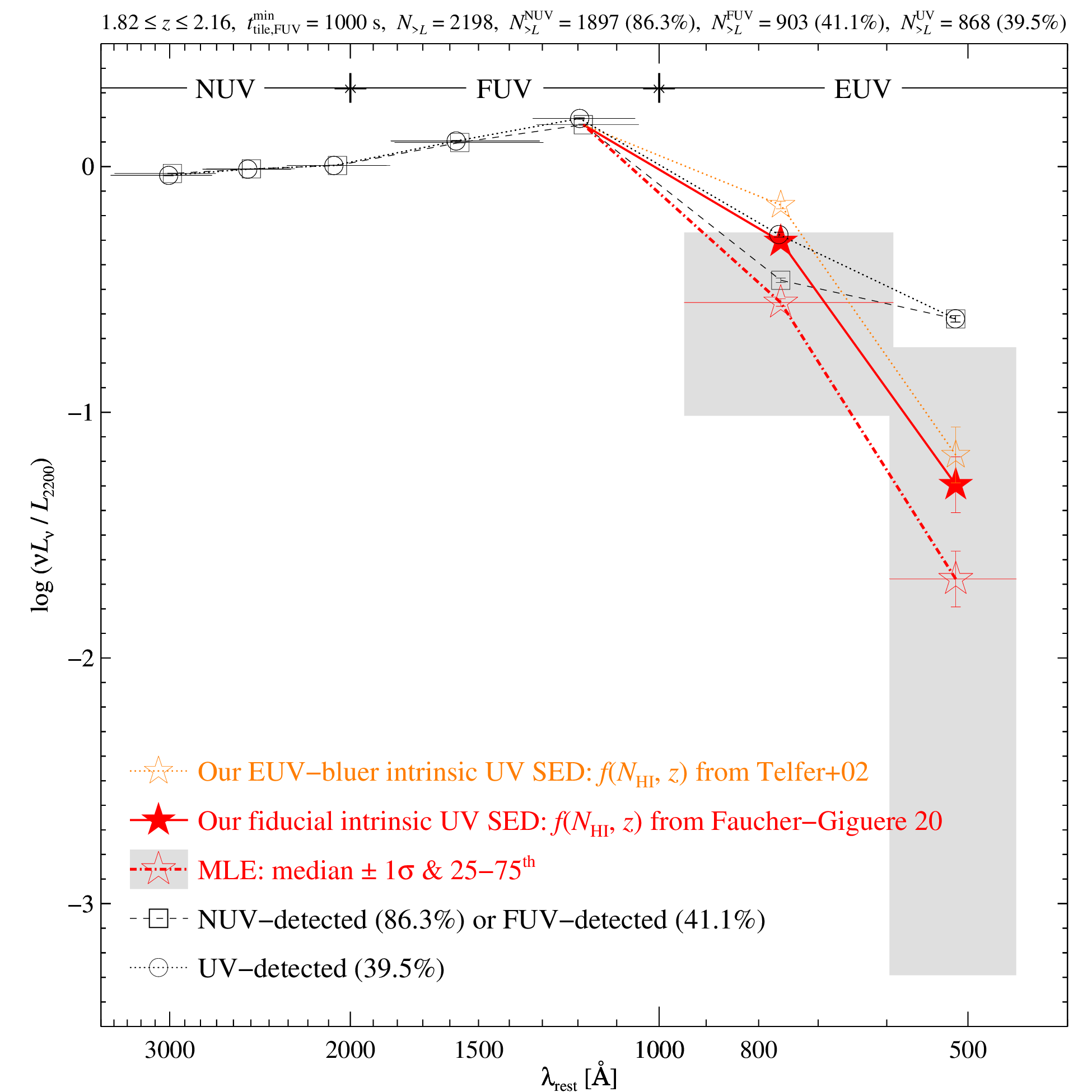}
\caption{
The median UV SEDs inferred from a bright quasar sub-sample with $46 < \log L_{2200} < 47.3$ and covered by \galex tiles whose FUV exposure times are longer than 1000 seconds. This sub-sample contains 2198 quasars; 86.3\% (41.1\%) of them are detected in NUV (FUV) and 39.5\% are detected in both UV bands. 
The median SED inferred from the $N^{\rm UV}_{>L}$ UV-detected sources (the black open circles linked by the dotted line) is globally consistent with that inferred from the whole sub-sample (the black squares linked by the dashed line; only detections in the observed-frame NUV/FUV band are treated), except in the observed-frame NUV band, where the latter has a factor of two higher NUV detection and so lower median SED.
Adopting the MLE method to account for the \galex upper detection limits, the median SED with a much redder/softer EUV shape is derived in the EUV part (the red open stars linked by the dot-dashed line superimposed on the light-gray regions for the 25-75th percentile range; see Figure~\ref{fig:dis_detected_upperlimit_NUV_FUV}).
Finally, correcting against IGM absorption gives rise to the intrinsic UV SEDs, assuming two different parameterizations for $f(N_{\rm HI}, z)$ from \citet[][the orange open stars linked by the dotted line for the EUV-bluer intrinsic UV SED]{Telfer2002a} and \citet[][the red filled stars linked by the solid line for the fiducial intrinsic UV SED]{Faucher-Giguere2020}.
}\label{fig:SED_EUV_intrinsic}
\end{figure*}

\subsubsection{A unique sub-sample of luminous quasars}

In previous studies, to correct for the incompleteness bias, we choose to match the detection completeness between various luminosity bins, which enables a fair comparison of SEDs between luminosity bins. However, the derived SED of {\it GALEX}-detected quasars only represents a small (practically with relatively harder SED) but not the whole population of quasars in each luminosity bin.
For a sample with extremely low \galex detection fraction (such as the lower luminosity bins in this work), it would be impossible to derive its intrinsic SED for the whole sample.

The right-panel of Figure~\ref{fig:sample_selection} shows that the UV detection fraction of quasars increases with increasing UV luminosity. Higher UV detection fraction is also expected for quasars covered by \galex tiles with longer exposure times. Therefore, by considering a sub-sample of luminous quasars with long \galex exposure such that the UV detection fraction is as large as possible, we may be able to derive a bias-free intrinsic mean EUV SED, by taking \galex non-detections into account through the survival analysis or fitting a parametrized distribution curve \citep{VandenBerk2020}. Since we have shown the SED exhibits no luminosity dependence, this bias-free mean SED derived from the most luminous sample could be used to represent the whole quasar population studied in this work.

Note that the number of quasars would significantly decreases with increasing luminosity and \galex exposure time. To help determining a proper criterion (including both luminosity and exposure time) to construct such a sub-sample with both relatively large UV detection fraction and sufficient quasars, Figure~\ref{fig:UV_detected_fractions_different_minimal_exposure_times_tFUV1_strict_longestFUV} shows, given three minimal \galex FUV exposure times, the quasar number (top panel), the NUV detection fraction (middle panel), and the FUV detection fraction (bottom panel) of quasars brighter than $\log L_{2200}$ as a function of $\log L_{2200}$. 
Table~\ref{tab:NUV_FUV_detection_limits} tabulates part of them.
Here, being different from what we treat above, we consider all detections in \galex catalog (without applying a further SNR cut) in NUV or FUV, in order to maximize \galex detections rather than adopting more uncertain upper limits (as described below) for sources with low SNR. We note that at $\log L_{2200} > 46$ or the FUV exposure time longer than 1000 seconds, the FUV detection fraction does not further increase much but the number of quasars decreases a lot. On balance between high NUV/FUV detection fraction and large quasar number, we determine to construct the bright sub-sample for quasars brighter than $\log L_{2200} = 46$ and covered by \galex tiles with a minimal FUV exposure time of 1000 seconds. In total, the bright sub-sample includes 2198 quasars with 86.3\% NUV and 41.1\% FUV detections (Table~\ref{tab:NUV_FUV_detection_limits}).

\subsubsection{The 3$\sigma$ \galex detection limits}

To derive the upper detection limits for those \galex non-detections, \citet{VandenBerk2020} have provided fits for the NUV/FUV limiting magnitudes as a function of the effective NUV/FUV exposure time (see their Equations 5-6). However, their upper detection limit is defined as ``the magnitude at which the number of detected objects per unit magnitude is half of the number expected to have been observed'', which means that indeed a significant fraction of \galex non-detected sources could be brighter than their 50\%-complete limiting magnitude.
Therefore, adopting such an upper limiting magnitude for those \galex non-detected quasars, one may underestimate \galex fluxes for a considerable fraction of them, making the derived mean EUV SED biased softer/redder.

To assess a more robust 3$\sigma$ NUV/FUV limiting magnitude, we plot in Figure~\ref{fig:detection_limits_NUV_FUV_texp_vb20} the NUV/FUV magnitudes versus exposure time for all NUV/FUV-detected DR14 quasars with \galex SNR of 2.9 -- 3.1 from the \citet{VandenBerk2020} catalog \footnote{\url{http://cdsarc.cds.unistra.fr/viz-bin/cat/J/MNRAS/493/2745}}.
We then derive the 3$\sigma$ limiting magnitude as a function of exposure time following \citet{VandenBerk2020}.
Clearly, the 3$\sigma$ limiting magnitudes are brighter (by $\simeq 0.71$ and $\simeq 0.41$ mag with 1$\sigma$ scatters of 0.34 and 0.33 mag in NUV and FUV, respectively) than the 50\%-complete limiting magnitudes given by \citet{VandenBerk2020}. 
We therefore assign each quasar without NUV/FUV detection a 3$\sigma$ NUV/FUV upper detection limit, according to the NUV/FUV exposure time of the \galex tile covering that quasar.
The upper detection limit assigned to a quasar is further adjusted by random Gaussian deviation to account for the clear scatter in the derived 3$\sigma$ limiting magnitudes from Figure~\ref{fig:detection_limits_NUV_FUV_texp_vb20}.

Relative to the rest-frame 2200~\angstrom~luminosity, distributions of the two EUV luminosities (corresponding to the observed-frame NUV and FUV bands) are shown in Figure~\ref{fig:dis_detected_upperlimit_NUV_FUV} for the said bright quasar sub-sample. 
Notably, the global distributions, consisting of both \galex detected and undetected quasars, show not only large dispersion but also significant asymmetry toward fainter luminosity. Much larger dispersion is found in the FUV band. The dispersion may be contributed by the intrinsic scatter of quasar EUV SEDs, such as variability, distinct geometry of the accretion flow, anisotropy of the disk emission, different inclination angle, and so on, but also could originate from a series of intrinsic absorptions by host galaxy, circumgalactic median, and IGM. The prominent asymmetry suggests important contribution by absorptions, especially in the FUV band.

To account for the upper limits and infer the relevant median/quartile values, we adopt both the survival analysis \citep[SA;][]{FeigelsonNelson1985} and the maximum-likelihood estimation \citep[MLE;][]{VandenBerk2020}. 
Adopting our 3$\sigma$ upper detection limits, we find that the median values derived through the two methods are consistent with each other and both are smaller than that implied by solely considering \galex detected quasars. 
The quartile values derived through the two methods are also consistent, except the lower quartile values in the FUV band (cf. the right-panel of Figure~\ref{fig:dis_detected_upperlimit_NUV_FUV}). The fact that the lower quartile value derived through the SA method is larger than the MLE method is because when estimating the potential NUV/FUV luminosities for the undetected quasars the SA method assumes a minimal luminosity value corresponding to the minimum of all upper detection limits, while the MLE method allows any arbitrarily values smaller than the detection limit.
Note, for this bright sub-sample, the NUV detection fraction is large enough ($\sim 86\%$) so the median values inferred in NUV band have rather small errors, estimated by bootstrapping. Instead, the FUV detection fraction is only $\sim 41\%$, resulting considerable uncertainties in the derived median values (which could be reduced with future deeper FUV imaging surveys, such as, to be conducted by UVEX; \citealt{Kulkarni2021}).

\subsubsection{Correction against IGM absorption}

To statistically correct the IGM absorption for our bright quasar sub-sample following the standard procedure \citep[e.g.,][]{MollerJakobsen1990,MeiksinMadau1993,MadauHaardt2009}, we model the IGM transmission as a result of absorption of both Lyman continuum (LyC) and Lyman series line (LyL) due to intervening Poisson-distributed neutral hydrogen (HI) clouds along the line of sight, such as Lyman forest systems and Lyman limit systems (with hydrogen column densities $N_{\rm HI}$ typically smaller and larger than $\sim 10^{17}~{\rm cm}^{-2}$, respectively).

The total IGM transmission is $T_\lambda = \exp(-\tau_{\rm eff})$, where the total effective optical depth, $\tau_{\rm eff} = \tau_{\rm eff}^{\rm LyC} + \tau_{\rm eff}^{\rm LyL}$, includes both the effective LyC optical depth, $\tau_{\rm eff}^{\rm LyC}$, and the effective LyL optical depth, $\tau_{\rm eff}^{\rm LyL}$. 
For a bunch of EUV photons emitted at $z_{\rm e}$ and with rest-frame wavelength $\lambda_{\rm e}$ smaller than Lyman limit $\lambda_{912} = 911.75$~\angstrom, they are subject to LyC absorption beyond $z_{912}$ by
\begin{align}
  \tau^{\rm LyC}_{\rm eff}(z_{\rm e}, \lambda_{\rm e}) &= \int^{z_{\rm e}}_{z_{912} \geqslant 0} dz \int^{N^{\rm max}_{\rm HI}}_{N^{\rm min}_{\rm HI}} d N_{\rm HI} \nonumber \\
  &\times f(N_{\rm HI}, z) \{ 1 - \exp[ - N_{\rm HI} \sigma(\lambda_{z}) ] \},
\end{align}
where $(1 + z_{\rm e}) \lambda_{\rm e} = (1 + z) \lambda_{z} = (1 + z_{912}) \lambda_{912}$, $f(N_{\rm HI}, z) = \partial^2 n/\partial z \partial N_{\rm HI}$ is the distribution of absorbers defined as the number of absorbers per unit redshift and per unit column density between $N_{\rm HI}^{\rm min}$ and $N_{\rm HI}^{\rm max}$, and $\sigma(\lambda_z) \simeq \sigma_{912} (\lambda_z/\lambda_{912})^3$ is the HI photoionization cross section with $\sigma_{912} = 6.35 \times 10^{-18}~{\rm cm}^2$.
Once these EUV photons are redshifted longer than $\lambda_{912}$ but smaller than Ly$\alpha$ wavelength $\lambda_{1216} = 1215.67$~\angstrom, they would be further undergone LyL absorption at a discrete set of redshifts according to 
\begin{align}
  \tau^{\rm LyL}_{\rm eff}(z_{\rm e}, \lambda_{\rm e}) &= \sum_{0 \leqslant z_n \leqslant z_{\rm e}} (1 + z_n) \int^{N^{\rm max}_{\rm HI}}_{N^{\rm min}_{\rm HI}} d N_{\rm HI} \nonumber \\
  & f(N_{\rm HI}, z_n) \frac{W_n[\tau_0(N_{\rm HI}, f_n), b, \gamma_n]}{\lambda_n},
\end{align}
where $(1 + z_{\rm e}) \lambda_{\rm e} = (1+z_n) \lambda_n $, $\lambda_n = \lambda_{1216} \frac{3}{4(1 - 1/n^2)}$ is the wavelength of the $1 s \rightarrow n p$ Lyman series transition ($n \geqslant 2$), and $W_n$ is the rest-frame equivalent width of the absorption line $\lambda_n$ in wavelength units. 
The optical depth at line-center $\tau_0 = \sqrt{\pi} \frac{e^2}{m_e c} f_n \lambda_n \frac{N_{\rm HI}}{b}$, where $e$ is the electron charge quantum, $m_e$ is the electron mass, $c$ is the speed of light, $f_n$ is the oscillator strength of the transition for line $\lambda_n$, and $b$ is the Doppler broadening parameter.
For the case in the IGM, the reciprocal mean lifetime of the $n$ state of neutral hydrogen, $\gamma_n$, is primarily attributed to the spontaneous decays, $A_{nj}$, from the $n$ state to all other lower levels $j$, that is, $\gamma_n \simeq \sum_{1 \leqslant j < n} A_{nj}$.
Adopting the approximation formulae for $W_n$ provided by \citet[][cf. their Equation~9.27]{Draine2010book}, we assume $b = 30~{\rm km~s^{-1}}$ and consider Lyman series lines up to $n = 20$, for which both $f_n$ and $A_{nj}$ are available in \citet{WieseFuhr2009}. 

The above assumptions for the LyL absorption would not significantly affect our IGM correction since the IGM transmission at $\lambda < \lambda_{912}$ where we are interested here is dominated by the LyC absorption. Instead, the LyC absorption is sensitive to $f(N_{\rm HI}, z)$ and the adopted dynamical range of $N_{\rm HI}$.
In the left panel of Figure~\ref{fig:IGM_transmission}, we show two parameterizations for $f(N_{\rm NI}, z)$, from \citet{Telfer2002a} and \citet{Faucher-Giguere2020}, respectively. The latter is an upgraded version based on \citet{HaardtMadau2012} and \citet{Puchwein2019} and shows a deficiency of absorbers with $N_{\rm HI} \sim 10^{17}-10^{19}~{\rm cm^{-2}}$, compared to that used by \citet{Telfer2002a}. 
Since the broadband rest-frame EUV emissions of our quasars spreading over a large sky area may be attenuated by absorbers with any column density, we adopt the whole range $12 \leqslant \log N_{\rm HI} \leqslant 22$ to estimate the mean IGM transmission for our quasars.
Considering $12 \leqslant \log N_{\rm HI} \leqslant 22$, higher IGM transmission fraction is implied by $f(N_{\rm NI}, z)$ from \citet{Faucher-Giguere2020} than that from \citet{Telfer2002a}, as illustrated in the right panel of Figure~\ref{fig:IGM_transmission}. However, we confirm that the IGM transmissions implied by $f(N_{\rm NI}, z)$ from \citet{Telfer2002a} and \citet{Faucher-Giguere2020} are comparable over the whole EUV range if adopting a smaller $N_{\rm HI}^{\rm max}$, that is, $12 \leqslant \log N_{\rm HI} \leqslant 16.7$. Over this range, $f(N_{\rm HI}, z)$ at $z = 2$ adopted in these two works are similar as shown in the left panel of Figure~\ref{fig:IGM_transmission}. Therefore, in the right panel of Figure~\ref{fig:IGM_transmission}, the lower IGM transmission implied by $f(N_{\rm NI}, z)$ from \citet{Telfer2002a} is primarily due to the overabundance of absorbers with $N_{\rm HI} \sim 10^{17}-10^{19}~{\rm cm^{-2}}$.

\subsubsection{The intrinsic mean UV SED}

Finally, as illustrated in Figure~\ref{fig:SED_EUV_intrinsic}, we show, step by step, how the derived median UV SED changes from solely considering the UV-detected quasars (the open circles linked by the dotted line), through including all NUV/FUV detections (the open squares linked by the dashed line), to being accounted for the NUV/FUV upper detection limits (the red open stars linked by the dot-dashed line), and then further being corrected against the IGM absorption (the red filled stars linked by the solid line versus the orange open stars linked by the dotted line).

First, the median UV SED for the UV-detected quasars (detected in both \galex NUV and FUV, only 39.5\%) is compared to that for the NUV- (86.3\%) or FUV-detected (41.1\%) quasars. The latter median UV SED (hereafter, the NUV/FUV-detected UV SED) is constructed for the whole sub-sample, except in the NUV/FUV band where only the NUV/FUV-detected quasars are considered. These two median UV SEDs are globally consistent, except in the observed-frame NUV band, where the latter has a factor of two higher NUV detection fraction and so lower median SED. In the observed-frame FUV band, the two SEDs are similar owing to their comparable detection fractions in that band. This also clearly implies that the derived SED in EUV is sensitive to the \galex detection fraction and one need to account for the NUV/FUV upper detection limit once pursuing the intrinsic one. 

Second, following \citet{VandenBerk2020}, we cross-check two methods to deal with the upper detection limit and find as shown in Figure~\ref{fig:dis_detected_upperlimit_NUV_FUV} that the median/quartile values implied by them are almost consistent, so for clarify we only show in Figure~\ref{fig:SED_EUV_intrinsic} the median EUV SED implied by the MLE method. Once accounting for the upper detection limits, the median EUV SED turns out to be much redder/softer, especially in the observed-frame \galex FUV band where the detection fraction is lower.

Third, using the two distinct IGM transmission curves shown in the right panel of Figure~\ref{fig:IGM_transmission}, we derive the intrinsic UV SEDs for our bright quasar sub-sample. We prefer the fiducial intrinsic UV SED implied by the up-to-date $f(N_{\rm HI}, z)$ from \citet{Faucher-Giguere2020}, but also present another EUV-bluer intrinsic UV SED implied by $f(N_{\rm HI}, z)$ from \citet{Telfer2002a} in order to highlight the underlying uncertainty of IGM correction.
{Note that the uncertainty of IGM correction has not been transported to all our intrinsic UV SEDs.}

\subsection{Comparison with previous results}\label{sect:previous_results}

\begin{figure*}[t!]
\centering
\includegraphics[width=0.8\textwidth]{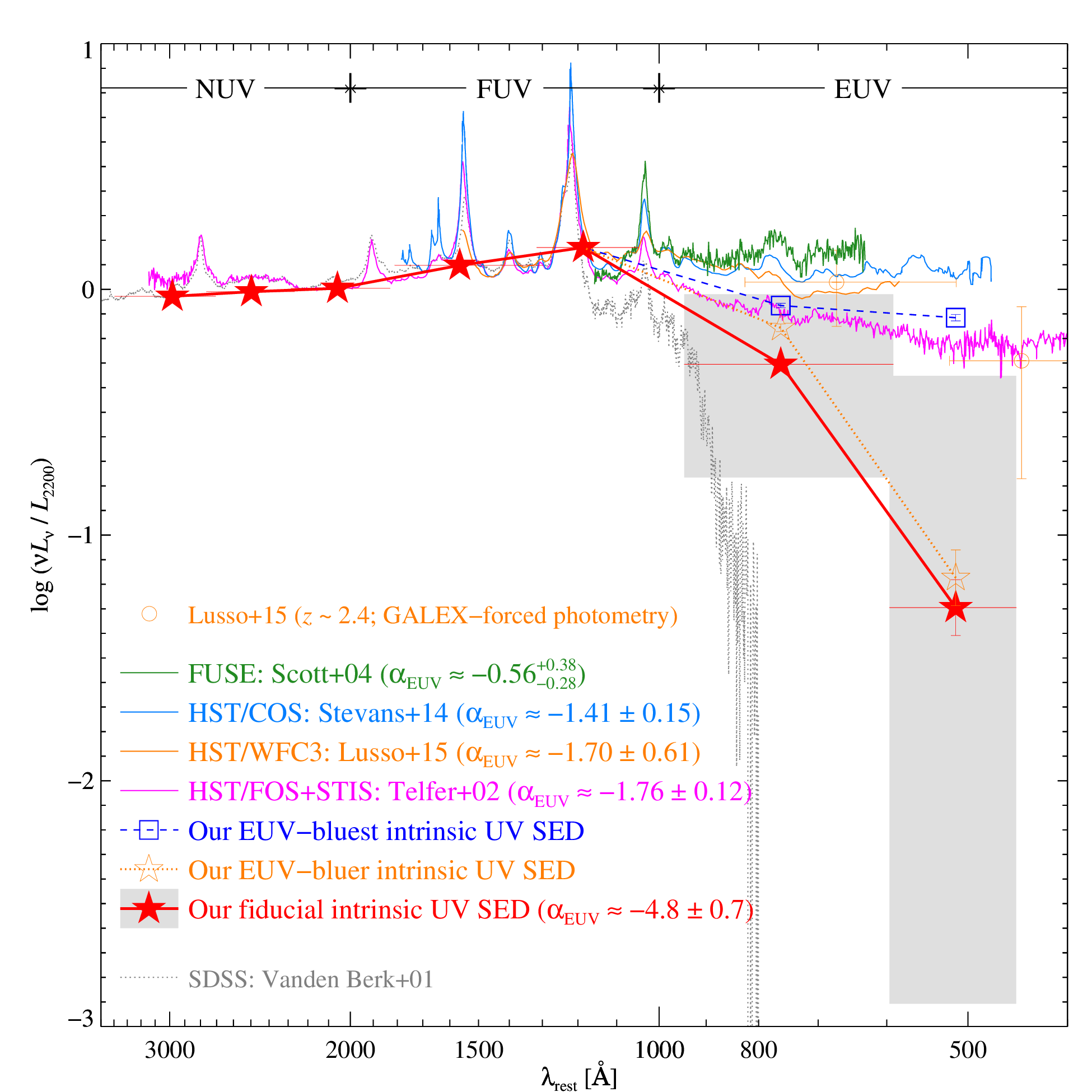}
\caption{
Our IGM-corrected intrinsic mean UV SEDs are compared to several previous composite quasar spectra.
Besides the fiducial (the red filled stars linked by the red thick solid line superimposed on the light-gray regions for the 25-75th percentile range in the EUV part; see Section~\ref{sect:intrinsic_mean_SED}) and EUV-bluer (the orange open stars linked by the dotted line) intrinsic UV SEDs, an EUV-bluest intrinsic UV SED (the blue open squares linked by the dashed line) is also illustrated by adopting $f(N_{\rm HI}, z)$ from \citet{Telfer2002a} to correct the NUV/FUV-detected UV SED, which indicates an upper limit for our intrinsic UV SED.
Except the SDSS quasar composite spectrum from \citet{VandenBerk2001}, the others have been applied corrections for Lyman absorptions. The composite spectra from \citet{VandenBerk2001} and \citet[][see \citealt{Zheng1997} for a similar composite spectrum]{Telfer2002a} are normalized at 2200~\angstrom. Owing to the limited wavelength coverage and following \citet{Lusso2015}, the composite spectra from \citet[][see \citealt{Shull2012} for a similar composite spectrum]{Stevans2014} and \citet[][the two orange open circles for their IGM-corrected mean {\it GALEX}-forced photometry]{Lusso2015} are normalized to that of \citet{Telfer2002a} at 1450~\angstrom, while the one from \citet{Scott2004} is normalized to that of \citet{Lusso2015} at 1114~\angstrom.
The EUV spectral indices $\alpha_{\rm EUV}$ at $\lambda_{\rm rest} < 1000$~\angstrom~are taken from the corresponding references, while our EUV index is estimated solely using the two \galex photometric points.
}\label{fig:SED_EUV_intrinsic_comp_spectroscopic}
\end{figure*}

There are many previous studies exploring the mean SED or composite spectrum of quasars in the rest-frame EUV, using either photometric \citep[e.g.,][]{Trammell2007,Krawczyk2013} or spectroscopic data \citep[e.g.,][]{Zheng1997,VandenBerk2001,Telfer2002a,Shull2012,Stevans2014,Lusso2015}. As we have discussed above, the shape/slope of the mean UV SED of a rest-frame EUV-detected sample severely depend on the EUV detection completeness which is sensitive to luminosity/redshift ranges, thus it would be infeasible to directly compare those SEDs from various photometric studies without correcting the sample incompleteness bias. 
While we have previously discussed other photometric studies \citep[e.g.,][]{Trammell2007,Krawczyk2013}, in this sub-section, we should only compare our intrinsic UV SEDs derived in Section~\ref{sect:intrinsic_mean_SED} to several composite quasar spectra, as illustrated in Figure~\ref{fig:SED_EUV_intrinsic_comp_spectroscopic}.

In Figure~\ref{fig:SED_EUV_intrinsic_comp_spectroscopic}, compared are several UV composite spectra of quasars corrected against Lyman absorption, except that of \citet{VandenBerk2001}, which is shown only to highlight the importance of applying correction for the IGM absorption. For those with the IGM correction, the EUV slopes are clearly diverse, probably indicating complicated and different sample selections plus IGM corrections, as already discussed by those references. Nevertheless, the difference of the EUV spectra between \citet{Telfer2002a} and \citet{Scott2004} has long been suggested as an indication of the luminosity dependence of the ionizing continuum, which in turn explains the Baldwin effect (however, see Section~\ref{sect:baldwin_effect} of this work).

On one hand, our fiducial UV SED (the red filled stars linked by the red thick solid line in Figure~\ref{fig:SED_EUV_intrinsic_comp_spectroscopic}) is much bluer that of \citet{VandenBerk2001} without IGM correction. Since the minimal observed wavelength of the SDSS spectrum is $\sim 3800$~\angstrom, the \citet{VandenBerk2001} composite spectrum at rest-frame $\sim 800$~\angstrom~is contributed by a few sources ($\lesssim 40$) at $z \gtrsim 3.7$ and therefore is subject to the IGM attenuation much stronger than our sample. Adopting $f(N_{\rm HI}, z)$ from \citet{Faucher-Giguere2020} and $12 \leqslant \log N_{\rm HI} \leqslant 22$, we estimate $T_\lambda(z_{\rm e} \gtrsim 3.7, \lambda_{\rm e} \sim 800~\angstroms) \lesssim 0.1$. Comparing the more than one dex difference at $\sim 800$~\angstrom~between their composite spectrum without IGM correction and our fiducial UV SED with IGM correction, if their composite spectrum were corrected against the IGM attenuation, it is likely consistent with our result.

On the other hand, our fiducial intrinsic UV SED is significantly redder than all other composite quasar spectra with IGM correction.
Using the two \galex photometry in the rest-frame EUV, we estimate our EUV spectral index to $\alpha_{\rm EUV} \simeq 4.8 \pm 0.7$, where the 1$\sigma$ uncertainty is estimated by bootstrapping the two photometry points independently. Our EUV spectral index significantly suggests a redder/softer mean EUV shape for quasars than the available reddest \citet{Telfer2002a} composite spectrum.
Even our EUV-bluer intrinsic UV SED (the orange open stars linked by the dotted line), which has enhanced the EUV emissions of quasars after IGM correction by overestimating the number density of absorbers with $N_{\rm HI} \sim 10^{17}-10^{19}~{\rm cm^{-2}}$, is still redder than all other composite quasar spectra.
Conservatively, we present in Figure~\ref{fig:SED_EUV_intrinsic_comp_spectroscopic} a special intrinsic UV SED for quasars with real NUV/FUV detections only (the blue open squares linked by the dashed line). Being complemented with the overestimated number density of absorbers, that is, using $f(N_{\rm HI}, z)$ from \citet{Telfer2002a}, this EUV-bluest intrinsic UV SED indicates a robust upper limit that can be derived from our quasar sample. Again, our EUV-bluest intrinsic UV SED is definitely redder than all other composite quasar spectra but the \citet{Telfer2002a} one. 

Actually, the completeness of quasar samples in works constructing composite quasar spectrum is hard to assess. These quasar samples spread over broad and diverse dynamical ranges of redshift and luminosity \citep[][cf. their Figure~9]{Lusso2015}. After applying several selection criteria, such as high spectral SNR, broad wavelength coverage, no abnormal emission/absorption lines, no prominent/suspected broad absorption lines, and no blazars, 
\citet{Scott2004} collect all {\it FUSE} spectra for 85 AGN at $0.01 < z < 0.67$ and with $43 \lesssim \log L_{1100} \lesssim 47$ as of 2002 November, 
\citet{Stevans2014} retrieve all {\it HST}/COS spectra for 159 AGN at $0.001 < z < 1.5$ and with $40 \lesssim \log L_{1100} \lesssim 46.7$ as of 2013 April, while 
\citet{Telfer2002a} compile all {\it HST}/FOS/GHRS/STIS spectra for 184 quasars at $0.33 < z < 3.7$ and with $44 \lesssim \log L_{1100} \lesssim 47.3$ as of 2000 August. 
Although all of these works have exhausted the corresponding archives, it does not necessarily guarantee that the completeness should be high since quasars targeted for UV spectroscopy are more likely kept in the archive if they are bluer in EUV, besides their selection function is extremely difficult to quantify \citep{Lusso2015}. Note that the broader dynamical range of UV luminosity covered by these quasar samples expectably brings about higher incompleteness (cf. the right-panel of Figure~\ref{fig:sample_selection}).
Contrariwise, \citet{Lusso2015} construct the composite spectrum for a quasar sample with much cleaner selection criteria. They employ the {\it HST}/WFC3 grism spectra for 53 quasars observed by \citet{OMeara2011} who select them from SDSS Data Release 5 \citep[DR5;][]{Schneider2007} with $g < 18.5$ mag and $2.3 < z < 2.6$. There 53 quasars have much narrower dynamical ranges of redshift and luminosity ($46.7 \lesssim \log L_{1350} \lesssim 47.2$) than the aforementioned quasar samples.
However, their sample completeness is still not clear enough. Using the same selection criteria, we find that there are indeed 316 quasars in SDSS DR5 \footnote{\url{http://classic.sdss.org/dr5/products/value_added/qsocat_dr5.html}} and so their completeness is likely only $\sim 17\% (=53/316)$.
To construct a specific quasar sample for studying the intervening Lyman limit system absorption, \citet{OMeara2011} further require no strong broad absorption line or $z \sim z_{\rm e}$ associated absorption line signatures in the SDSS quasar spectrum. Then, their parent quasar number may decrease to 100 \citep[cf. Table 1 of ][]{OMeara2011} and the completeness may increase to an upper limit of 53\%$(=53/100)$. For the latter parent quasars, distributions of redshift and luminosity for quasars observed with WFC3 are not different to those without.
However, no more information on how do they exclude quasars with absorption line limits a validation for the exact completeness and the selection function of the \citet{Lusso2015} sample.

Attractively, our bright quasar sub-sample even with only two EUV photometry points contains numerous quasars (2198) and has well-defined completeness (Section~\ref{sect:intrinsic_mean_SED}). Moreover, our detection fraction at $\sim 800$~\angstrom~is as high as 86.3\%.
For comparison, the numbers of quasars contributed to the composite spectrum at $\sim 800$~\angstrom~are only $\sim 10$, $\sim 50$, 53, and $\sim 60$ for \citet[][cf. their Figure~1]{Scott2004}, \citet[][cf. their Figure~2]{Stevans2014}, \citet[][]{Lusso2015}, and \citet[][cf. their Figure~2]{Telfer2002a}, respectively. 
The fact that the trend of increasing quasar number coincides with that of getting EUV slope redder from \citet{Scott2004} to \citet{Telfer2002a} may hint an increasing higher completeness.
Therefore, we are confident that our fiducial intrinsic UV SED at $\sim 800$~\angstrom~is more robust than all of them.

At the rest-frame $\sim 500$~\angstrom, our fiducial intrinsic UV SED is significantly redder than all previous composite spectra. We caution that our result at $\sim 500$~\angstrom~may be a bit uncertain since there is only 41.1\% real detections. However, the number of our real detected quasars is 903, which is already sufficiently lager than 1 (up to 4) and 8 quasars contributing to $\sim 500$~\angstrom~in \citet{Scott2004} and \citet{Telfer2002a}, respectively. 
We then suggest our fiducial intrinsic UV SED at $\sim 500$~\angstrom~is hitherto the most robust as well.
On the other hand, the extremely large dispersion of our quasar SEDs at $\sim 500$~\angstrom~may help alleviating the tension between our fiducial intrinsic UV SED and all other composite spectra since a large portion ($\lesssim 25\%$) of our quasars have EUV spectral shape quiet consistent with the quasar composite spectra.

Note that quasars with prominent broad absorption lines are generally excluded by works constructing the composite quasar spectrum. By checking the ${\rm BAL\_FLAG}$ or requiring the \CIV~absorption troughs wider than 2000 ${\rm km~s^{-1}}$ in the \citet{Rakshit2020} catalog, we find that both our parent SDSS quasar sample and the UV-detected one includes only $\sim 5\%$ broad absorption line quasars. After excluding these broad absorption line quasars, we confirm that our conclusions are preserved.

In sum, the previous quasar composite spectra in the EUV may have large uncertainties and be biased because of both the unknown selection function and the few number of individual spectra used in constructing the composite spectrum. Our analysis highlights the significance of properly considering the sample completeness before getting firm conclusions on the EUV property of quasars, such as their luminosity dependence. In the future, a large and complete SDSS quasar sample in a narrow high redshift range such that the EUV spectral coverage is reachable, say the CSST slitless spectroscopy \citep{Zhan2018,Zhan2021}, will be extremely helpful on addressing these issues.

\subsection{Physical implications by the universal mean UV SED}\label{sect:physical_implication}

\begin{figure*}[th!]
\centering
\includegraphics[width=0.33\textwidth]{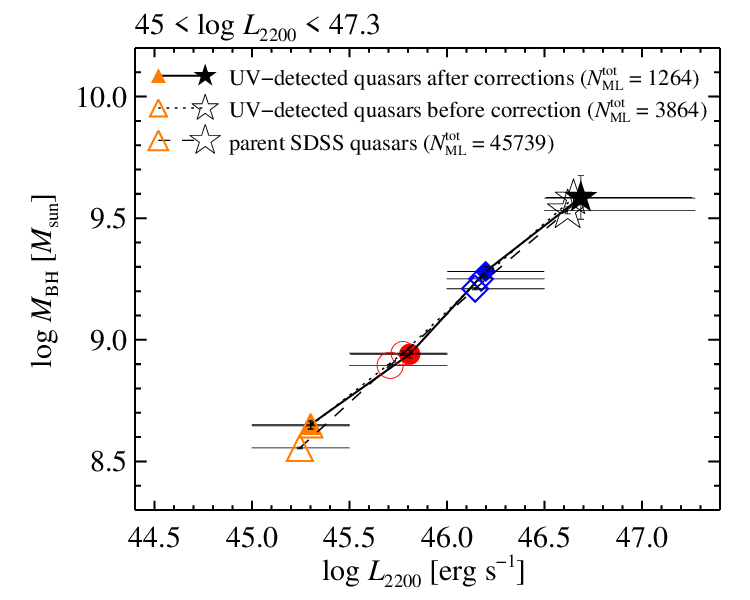}
\includegraphics[width=0.33\textwidth]{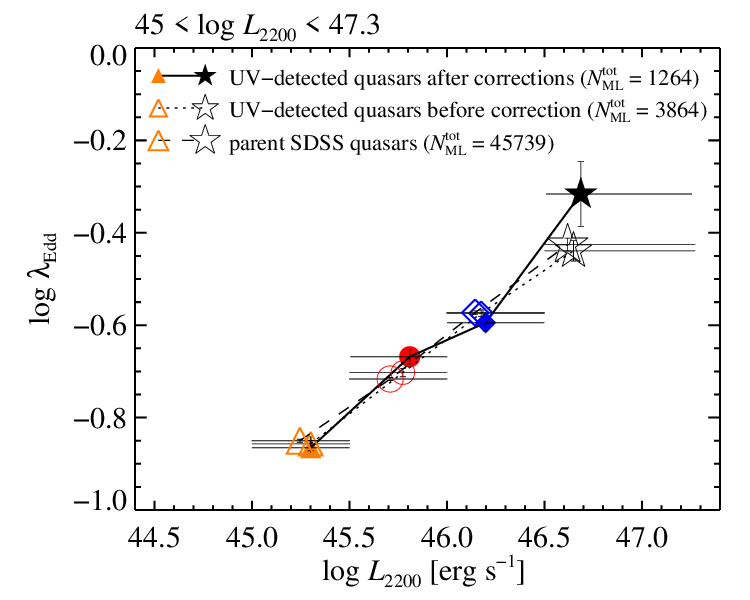}
\includegraphics[width=0.33\textwidth]{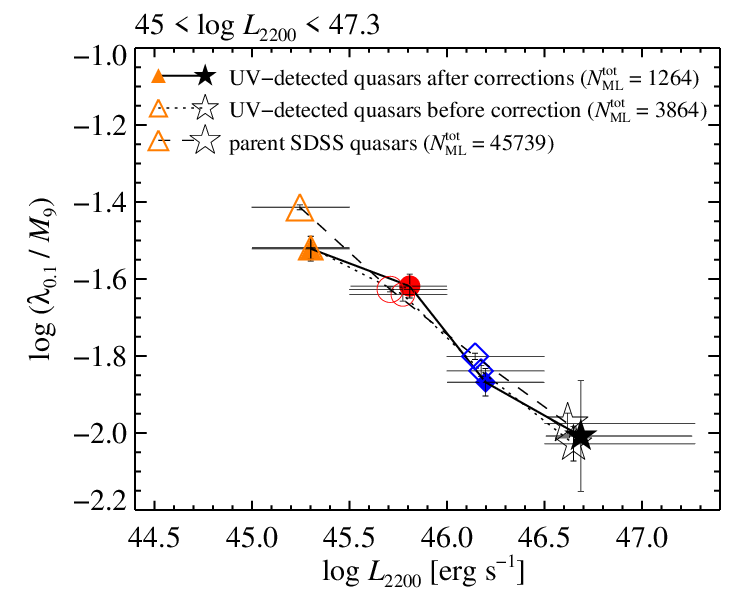}
\caption{ 
From left to right panels, the typical $M_{\rm BH}$, $\lambda_{\rm Edd}$, and $\lambda_{\rm Edd}/M_{\rm BH}$ as a function of 2200~\angstrom~luminosity ($45 < \log L_{2200} < 47.3$) are illustrated for quasars with available measurements of $M_{\rm BH}$ and $\lambda_{\rm Edd}$. Three sub-samples are compared: the parent quasar sample (large open symbols linked by the dashed lines), the UV-detected quasars before our bias correction (small open symbols linked by the dotted lines), and the UV-detected quasars after our bias correction (small solid symbols linked by the solid lines; see Section~\ref{sect:uniform_sed}). In each sub-sample, the total number of quasars with measurements of $M_{\rm BH}$ and $\lambda_{\rm Edd}$ is nominated as $N^{\rm tot}_{\rm ML}$.
In each luminosity bin, the median with minimum to maximum of 2200~\angstrom~luminosity of quasars are shown, while the median of each physical quantity of quasars is attached with $1\sigma$ error estimated as the standard deviation divided by the square root of the number of quasars in the corresponding luminosity bin.
Note in the right panel $M_9$ and $\lambda_{0.1}$ are $M_{\rm BH}$ and $\lambda_{\rm Edd}$ normalized to 10$^9~M_\sun$ and 0.1, respectively. 
}\label{fig:physical_properties_corrected_logL45}
\end{figure*}

\begin{figure*}[th!]
\centering
\includegraphics[width=0.8\textwidth]{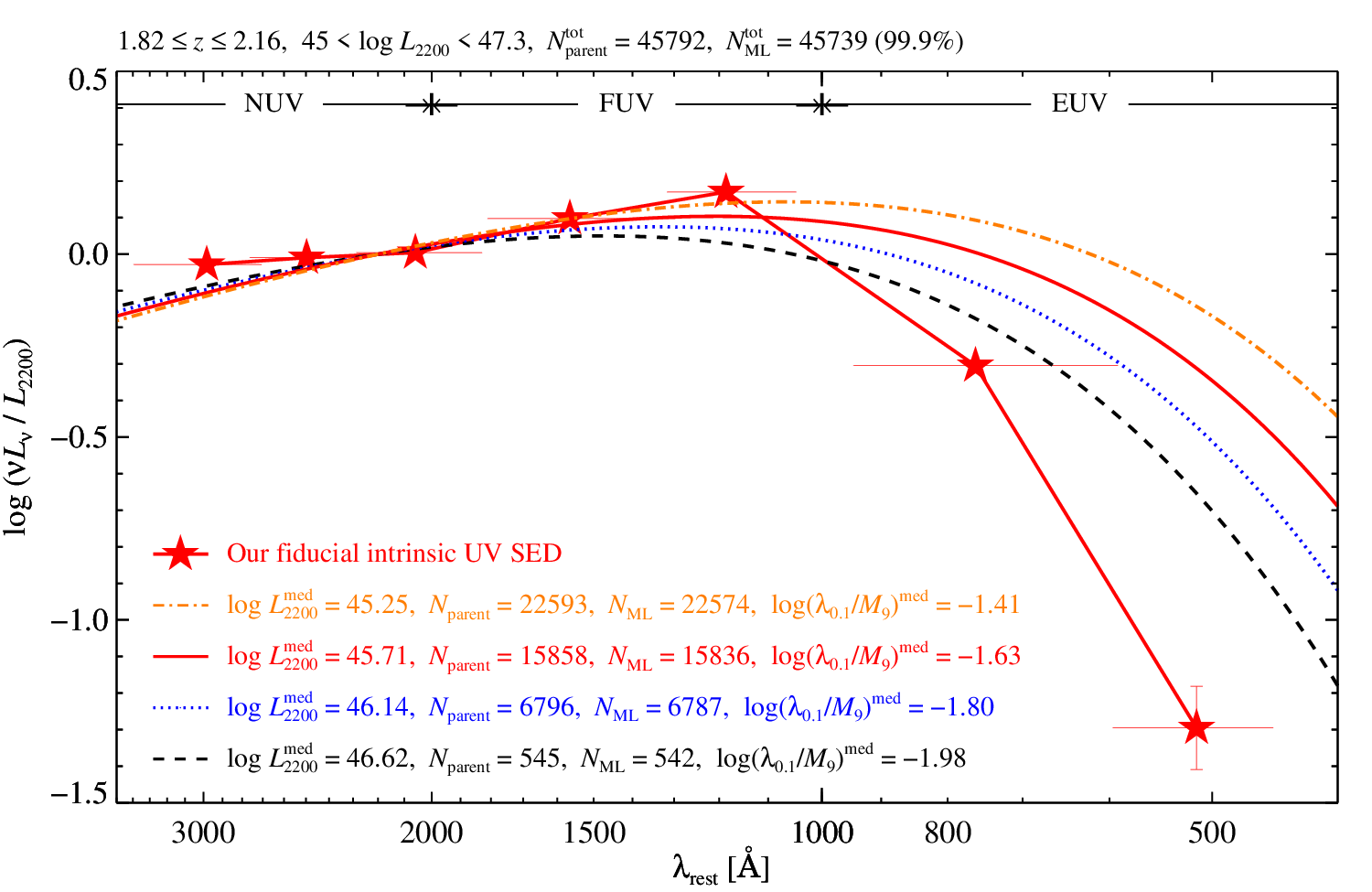}
\includegraphics[width=0.8\textwidth]{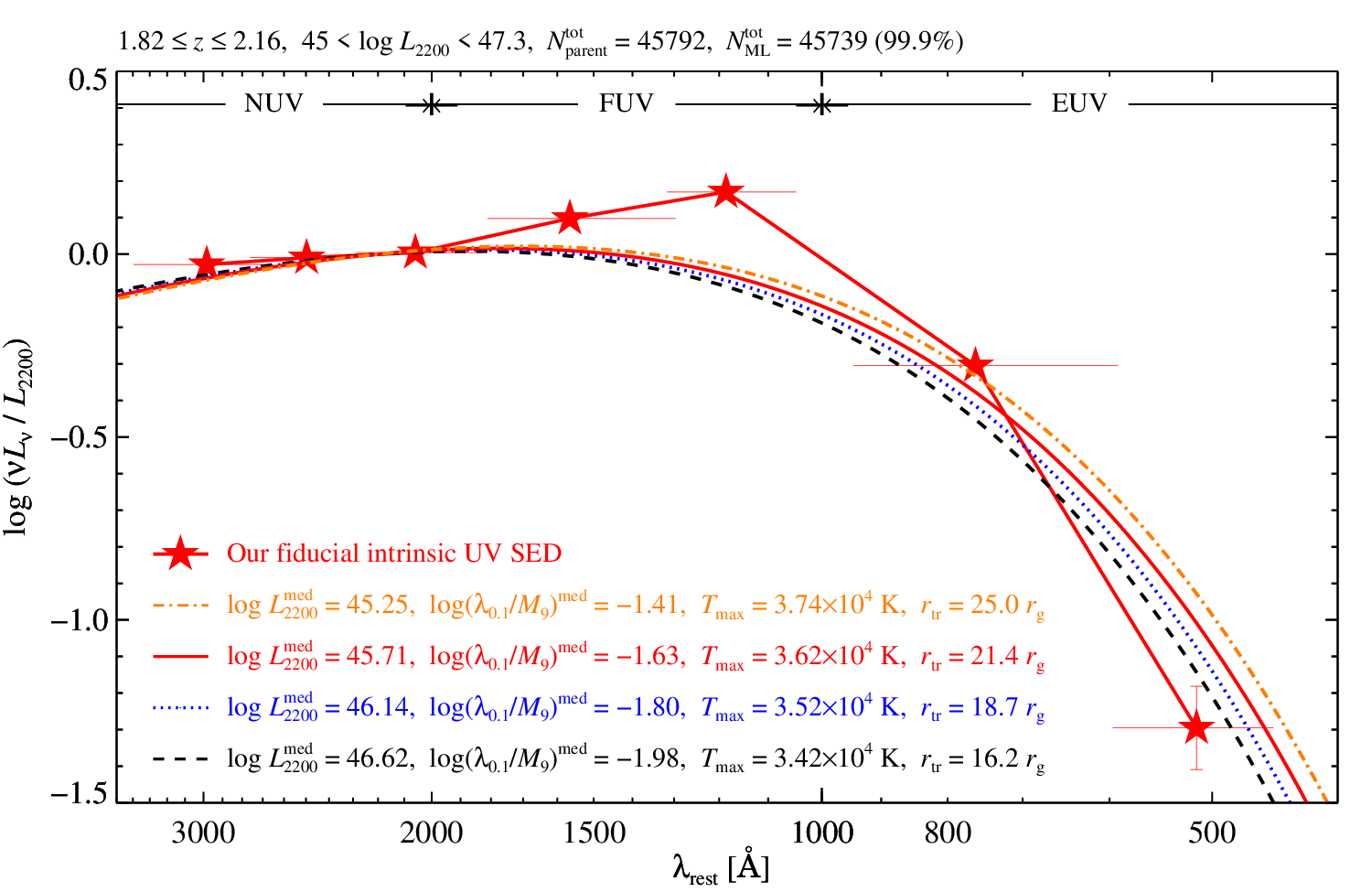}
\caption{
Top panel: for quasars with $45 < \log L_{2200} < 47.3$, the theoretical SEDs implied by a relativistic cold thin disk extending to $r_{\rm ISCO}$ of a non-rotating BH are compared to our fiducial intrinsic UV SED (the red filled stars linked by the red thick solid line; see Section~\ref{sect:intrinsic_mean_SED} and Figure~\ref{fig:SED_EUV_intrinsic}).
For the parent quasar sample in each luminosity bin, the median value of the 2200~\angstrom~luminosity, $\log L^{\rm med}_{2200}$, the number of all quasars, $N_{\rm parent}$, the number of those with available measurements on both BH mass and Eddington ratio, $N_{\rm ML}$, and the median value of the ratio of Eddington ratio to BH mass, $\log(\lambda_{\rm Edd}/M_{\rm BH})^{\rm med}$ (Figure~\ref{fig:physical_properties_corrected_logL45}), are nominated from left to right in the corresponding legend.
Bottom panel: same as the top one, but the theoretical SEDs are implied by the truncated cold thin disk model of \citet{LaorDavis2014}, where the truncated radius corresponds to $T_{\rm max}$ given by Equation~(\ref{eq:radius_t_max}). Each legend also includes $T_{\rm max}$ and the truncated radius $r_{\rm tr}$.
}\label{fig:SED_intrinsic_model_thindisk}
\end{figure*}

Before discussing the physical implications of the potential universal mean UV SED for quasars with $\log L_{2200} > 45$ on the central engine of AGN, we examine the difference of the BH mass and Eddington ratio among quasars in distinct luminosity bins. 
\citet{Rakshit2020} have estimated the BH mass and Eddington ratio for nearly all DR14Q quasars. Adopting their fiducial estimates, we find that both BH mass and Eddington ratio typically increase with UV luminosity, 
while $\lambda_{\rm Edd}/M_{\rm BH}$ shows the opposite trend (Figure~\ref{fig:physical_properties_corrected_logL45}). This is true not only for the parent quasar sample, but also for the UV-detected quasar samples before and after our bias correction introduced in Section~\ref{sect:uniform_sed}. The global consistence of these physical quantities between our UV-detected samples and the parent sample suggest that they are representative of the parent sample and the luminosity independence of the mean UV SED concluded by the UV-detected sample would also hold for the parent sample. 
Then, as illustrated in Figure~\ref{fig:SED_intrinsic_model_thindisk}, we would be able to compare the theoretical SEDs implied by models of the accretion flow adopting the typical $\lambda_{\rm Edd}/M_{\rm BH}$ of the parent sample in each luminosity bin to our intrinsic mean UV SED presented in Section~\ref{sect:intrinsic_mean_SED}.

The ratio of $\lambda_{\rm Edd}$ to $M_{\rm BH}$ uniquely determines the shape of SED for a pure cold thin disk which extends down to the innermost stable circular orbit radius, $r_{\rm ISCO}$, surrounding a non-rotating SMBH \citep[e.g.,][]{KoratkarBlaes1999}. Assuming a relativistic cold thin disk extending to $r_{\rm ISCO}$ of a Schwarzschild BH \citep{NovikovThorne1973} and adopting the typical $\lambda_{\rm Edd}/M_{\rm BH}$ for each luminosity bin, the top panel of Figure~\ref{fig:SED_intrinsic_model_thindisk} shows the resultant multi-color blackbody SED normalized at 2200~\angstrom. Clearly, the pure cold thin disk model using the observed typical BH mass and Eddington ratio indeed predicts luminosity-dependent EUV SEDs (bluer for less luminous ones) for our quasars. Contrarily, our analyses in previous sections suggest a universal mean UV SED, which in turn suggests some properties of the accretion flow in quasars should be luminosity independent and be self-regulated according to some specific physics.

To account for the unique turnover at $\sim 1000$~\angstrom~in the mean AGN UV SED, \citet{LaorDavis2014} propose a line-driven wind model where the cold thin disk truncates at around a maximum temperature of 
\begin{equation}\label{eq:radius_t_max}
    T_{\rm max} \simeq 3.58 \times 10^4 \left( \frac{\lambda_{0.1}/M_{9}}{0.02} \right)^{0.07}~{\rm K},
\end{equation}
where $\lambda_{0.1} = \lambda_{\rm Edd}/0.1$ and $M_9 = M_{\rm BH} / 10^9 M_\sun$.
When disk temperature approaches to that, strong wind surges. They predict the cold thin disk should be truncated at a few tens of gravitational radius ($r_{\rm g} \equiv G M_{\rm BH} / c^2$), as a result of the balance between mass-accretion and mass-loss, and so the turnover at $\sim 1000$~\angstrom~is weakly dependent on the BH mass and AGN luminosity. For instance, our fainter quasars have typically larger $\lambda_{\rm Edd}/M_{\rm BH}$, or equivalently, larger temperature and stronger disk wind at the same radius, thus larger truncated radius is expected according to their model.
As illustrated in the bottom panel of Figure~\ref{fig:SED_intrinsic_model_thindisk}, weakly luminosity-dependent UV SEDs are indeed implied by a simple truncated cold thin disk whose truncated radius, $r_{\rm tr}$, corresponds to $T_{\rm max}$ given by Equation~(\ref{eq:radius_t_max}) adopting the typical $\lambda_{\rm Edd}/M_{\rm BH}$ of the parent sample in each luminosity bin. 
Strikingly, the implied UV SEDs are also almost consistent with our intrinsic universal mean UV SED. The data excess at around $1000 - 2000$~\angstrom~is likely due to the prominent \Lya~and \CIV~line emissions from the broad line region rather than the disk. This global consistence between the wind model and our data would then provide another evidence for the prevalent winds from the accretion disk in quasars and the alleviated BH growth  \citep[e.g.,][]{KingAR2010,SloneNetzer2012,LaorDavis2014}.
It is worth mentioning that the wind model proposed by \citet{Sun2019a} could also resolve the disk size tension between the thin disk prediction and that constrained by the microlensing observations \citep[e.g.,][]{Morgan2010}. 
If the wind is further the origin of clouds in the broad line region, disks in fainter AGN are likely launching more clouds since they are truncated at larger radii with shallower gravitational potential. This again supports the main origin of the Baldwin effect is not dominated by the ionizing continuum as we discussed in Section~\ref{sect:baldwin_effect}. In this case, the local disk turbulence may help launching wind easier, but a new wind model with disk turbulence would be required to simultaneously account for properties of both the UV SED and the AGN variability.

Compared to the weakly luminosity-dependent EUV SEDs shown in the bottom panel of Figure~\ref{fig:SED_intrinsic_model_thindisk}, an exact luminosity-independent UV SED can be achieved by slightly adjusting the truncated radius since there are uncertainties of both the measured $\lambda_{\rm Edd}/M_{\rm BH}$ and the assumed crude truncated disk model. On one hand, at $r > r_{\rm tr}$ the mass accretion rate should have decreased with decreasing radius, thus redder EUV SED is expected than that implied by sharply terminating the accretion at $r_{\rm tr}$. On the other hand, bluer EUV SED would be expected once taking into account the increasing opacity of electron scattering, and so the enhancing Comptonization of the thermal disk emission, in the disk atmosphere with increasing temperature at $T > 10^4$ K \citep{CzernyElvis1987,Done2012}. These two effects may counterbalance and again result in an EUV SED similar to that implied by the current simplified treatment, although a more realistic truncated disk model is required.

Meanwhile, 
\citet{CzernyElvis1987} suggest that the universality of AGN SED peaking at $\sim 1000$~\angstrom~is probably attributed to the onset of electron scattering, a local atomic-originated process in the disk atmosphere. An alike atomic process has also been discussed by \citet{Lawrence2012} and \citet{LaorDavis2014}. The same atomic process may also account for our universal mean EUV SED. However, a delicate comparison is out of the scope of this paper for two main reasons.
One reason on observational point of view is that our universal mean EUV SED introduced in Section~\ref{sect:intrinsic_mean_SED} has not been corrected for the intrinsic extinction/absorption due to the dust and neutral hydrogen both in the host galaxy and the circumnuclear region of quasar, and even in the circumgalactic median \citep{Rudie2013}. If the intrinsic extinction could be corrected, the mean EUV SED is expected to be bluer. Here we expect that the final mean EUV SED is also universal since the distributions of the rest-frame NUV/FUV colors are indistinguishable among different luminosity bins (the lower-two right panels of Figure~\ref{fig:dis_z_EBV_color_before_after_corrections} and Section~\ref{sect:uniform_sed}). 
The other reason on theoretical landscape is that an exact EUV SED is also determined by the detailed treatments of the Comptonization in the disk atmosphere \citep{CzernyElvis1987,Hubeny2001}, the disk vertical structure \citep{Hubeny1998}, and even the disk turbulence \citep{DexterAgol2011,Cai2016}. It is also likely in close relationship with the unclear soft X-ray excess and even the X-ray emissions \citep{Done2012,Petrucci2013,KubotaDone2018,HickoxAlexander2018}. 

Examining either ROSAT or {\it XMM-Newton} detection for all DR14Q quasars, only 22,989 ($\simeq 4.4\%$) are X-ray-detected \citep{Paris2018}. Both our parent and UV-detected quasar samples have comparable small X-ray detection fractions. Therefore, it is currently unreliable to extend our SED analysis to the X-ray. But the all sky X-ray survey being conducted by eROSITA about 25 times more sensitive than ROSAT will provide millions of AGN and will definitely revolutionize our understanding on the most energetic inner emissions of AGN \citep{Predehl2021}.
Another prospect unveiling the accretion physics is to explore the luminosity dependence of the EUV SED for quasars fainter than analyzed here, which would rely on future surveys deeper than SDSS to overcome its incompleteness at $\log L_{2200} < 45$ (Figure~\ref{fig:sample_selection}). 
Again, exploring either the shorter wavelength or the fainter luminosity, the detection incompleteness highlighted in this work should be carefully considered.

Last but not least, although both our luminosity-independent universal mean UV SED and the global consistence between our intrinsic mean UV SED and the wind model all suggest an origin of local physical process, we note there are large scatter of the EUV SED slope (Figures~\ref{fig:SED_EUV_intrinsic_comp_spectroscopic}), which is likely attributed to the intrinsic scatter of quasar EUV properties and the intrinsic absorptions (Section~\ref{sect:intrinsic_mean_SED}). Once the intrinsic absorptions are controlled in future works, properties of the EUV scatter and its luminosity dependence are expected to tell us more physics about the accretion flow.

\section{Conclusions}\label{sect:conclusions}

Utilizing the SDSS DR14Q and the final \galex GR6/7 catalogs, we have compiled a unique quasar sample at $1.82 \lesssim z \lesssim 2.16$ (Section~\ref{sect:sample_summary}) to explore their mean rest-frame UV to EUV SED and the accretion physics of the innermost disk regions.
After having carefully corrected against a severe observational bias due to the UV detection incompleteness, i.e., the more luminous in observed-frame optical, the more likely detected in UV (Section~\ref{sect:perform_corrections}), we find that the rest-frame $\sim$ 500 -- 3000 \angstrom\ SED exhibits no luminosity dependence (at $\log L_{2200} \simeq 45$ -- 47.3, Sections~\ref{sect:uniform_sed} and \ref{sect:different_matching_strategy}).
The luminosity independence of the SED implies the Baldwin effect observed in this luminosity range (Section~\ref{sect:baldwin_effect}) is not driven by SED slope, but the properties of the broad emission line region which could be associated with luminosity-dependent disk turbulence as recently suggested by \citet{Kang2021}.

Taking the \galex non-detections into account and further correcting against the IGM absorption, a bias-free intrinsic $\sim$ 500 -- 3000 \angstrom\ SED is derived from a bright quasar sub-sample (Section~\ref{sect:intrinsic_mean_SED}). This bias-free mean SED is found to be significantly redder in the EUV than all available composite quasar spectra reported in literature, and the discrepancy is likely due to much smaller samples utilized to derive the composite spectra and the associated sample incompleteness having not been properly accounted for  (Section~\ref{sect:previous_results}).
Both the extremely red EUV SED and its luminosity independence severely challenge the standard cold thin disk model (Section~\ref{sect:physical_implication}). 
While there clearly are other factors which could significantly affect the UV - EUV SED of quasars, including attenuation intrinsic to the quasar/host galaxy, disc wind/outflow, disk turbulence, X-ray corona, and so on, the universal (luminosity-independent) $\sim$ 500 -- 3000 \angstrom\ SED and its large scatter presented in this work could be used in the future to testify these possibilities. For instance and interestingly,  
the global consistence of our observed SED with a line-driven wind model \citep{LaorDavis2014} suggests that the universal mean EUV SED could be the result of a local atomic-originated process, that is the onset of electron scattering beyond $10^4$ K.

Cautions to the sample incompleteness, especially cross-matching sample among multi-wavelengths, should be taken seriously. Complete quasar samples from future surveys, such as combining WFST/LSST with CSST and eROSITA, are required to extend the SED studies to lower luminosity regime and the X-ray band, before shedding light on the yet unclear accretion physics of the innermost disk regions in AGN.

\acknowledgments

Z.Y.C. is grateful to Fei-Fan Zhu for his instruction in utilizing the \galex database and Jia-Lai Kang for help on the survival analysis.
This work is supported by the National Key R\&D Program of China No.2022YFF0503402, and the National Science Foundation of China (grant Nos. 12033006, 11890693, 12192221, and 11873045). Z.Y.C. acknowledges support from the USTC Research Funds of the Double First-Class Initiative with No. YD2030002009, the science research grants from the China Manned Space Project with No. CMS-CSST-2021-A06, and the Cyrus Chun Ying Tang Foundations.
\vspace{5mm}

\bibliographystyle{aasjournal}
\bibliography{uvsed_230901_arxiv}

\end{document}